\documentclass[%
 reprint,
nofootinbib,
amsmath,amssymb,
aps,
prd,
]{revtex4-2}

\usepackage{aas_macros}
\usepackage{graphicx}
\usepackage{dcolumn}
\usepackage{bm}
\usepackage{xcolor}
\usepackage{amssymb,amsfonts}
\usepackage{graphicx,longtable,mathrsfs,color,array}
\usepackage{hyperref}
\usepackage{booktabs}
\usepackage{placeins}
\usepackage{bm}
\usepackage{siunitx}
\usepackage{bold-extra}
\usepackage{soul}

\newcommand{\comment}[1]{}
\newcommand{\cgw}{c_{\rm gw}}

\begin{document}

\preprint{APS/123-QED}

\title{Testing general relativity with gravitational waves---improving and extending Modified Dispersion Relation tests}

\author{Tomasz Baka$^{1,2}$}
\author{Bal\'azs Cirok$^3$}
\author{K Haris$^4$}
\author{Johannes Noller$^{5,6}$}
\author{N. V. Krishnendu$^{7,8}$}

\affiliation{${}^1$Institute for Gravitational and Subatomic Physics (GRASP), Utrecht University, 3584 CC Utrecht, The Netherlands}
\affiliation{${}^2$Nikhef, 1098 XG Amsterdam, The Netherlands}
\affiliation{${}^3$Department of Theoretical Physics, University of Szeged, Tisza Lajos krt. 84-86, H-6720 Szeged, Hungary}
\affiliation{${}^4$Department of Physics, National Institute of Technology, Kozhikode, Kerala 673601, India}%
\affiliation{${}^5$Department of Physics \& Astronomy, University College London, London, WC1E 6BT, U.K.}
\affiliation{${}^6$Institute of Cosmology \& Gravitation, University of Portsmouth, Portsmouth, PO1 3FX, U.K.}
\affiliation{${}^7$School of Physics and Astronomy, University of Birmingham, Edgbaston, Birmingham, B15 2TT, UK}
\affiliation{${}^8$International Centre for Theoretical Sciences (ICTS), Survey No. 151, Shivakote, Hesaraghatta, Uttarahalli, Bengaluru, 560089, India}

\date{\today}

\begin{abstract}
Searching for a modified dispersion relation is one of the general relativity tests performed by the LIGO-Virgo-KAGRA collaboration with each new cumulative Gravitational Wave Transient Catalog (GWTC). It considers classes of theories that modify the dispersion of gravitational waves by introducing a massive graviton or breaking Lorentz invariance. The symmetry breaking is parameterized phenomenologically by a momentum power law term $p^\alpha$ added to the dispersion relation, with the test placing constraints on the amplitude $A_\alpha$ of the introduced deviation. In this work, we implement improvements to the test, chief among them group velocity parametrization, a better sampling procedure, and extension to negative exponents $\alpha$ of $p$. We then reanalyze the events from the third catalog, GWTC-3, with our improved method.
Compared with GWTC-3 results, the individual event posteriors have significantly higher effective sample size and fewer multimodalities. The combined posteriors on the amplitude parameters $A_\alpha$ are, on average, 19\% narrower. The 90\% upper bound on the graviton mass changes from \num{2.42e-11} peV to \num{2.21e-11} peV. For the extension of our test to $\alpha \in \{-1, -2, -3\}$, we find no evidence in favor of general relativity violation.
\end{abstract}

\maketitle


\section{\label{sec:intro}Introduction}
\subsection{\label{sec:mdr-explanation}Search for modified dispersion relation}

The detection of gravitational waves (GWs) by the LIGO-Virgo-KAGRA (LVK) collaboration~\cite{2016PhRvL.116f1102A} marked a turning point in astronomy, offering a new way to study the universe~\cite{KAGRA:2013rdx, LIGOScientific:2014pky, VIRGO:2014yos, PhysRevD.88.043007} through the ripples in spacetime. 
These observations, particularly those from merging black holes and neutron stars, provided unprecedented opportunities to test Einstein's general theory of relativity (GR) with great precision on the data from GWTC-1-3 catalogs~\cite{2019PhRvD.100j4036A,2021PhRvD.103l2002A,2021arXiv211206861T}. While GR has consistently held up under scrutiny, researchers are always searching for potential deviations. One approach involves utilizing alternative theories of gravity, comparing the predicted gravitational waveforms from these models with the actual data collected by GW detectors. However, explicitly calculating these waveforms from alternative theories can be a complex and computationally challenging task.

Therefore, researchers often turn to theory-agnostic parameterized studies. Here, they do not rely on specific theories but instead quantify deviations from GR through effective measurable parameters~\cite{2021arXiv211206861T,2014PhRvD..89h2001A,2023PhRvD.107d4020M,2012PhRvD..85b4041M,2017PhRvL.119i1101K,2023PhRvD.108b4043M,2006CQGra..23L..37A,2009PhRvD..80l2003Y}.
For instance, there are many alternative theories of gravity predicting modifications to the dispersion relation of GWs~\cite{Calcagni:2009kc,AmelinoCamelia:2002wr, Horava:2009uw,Sefiedgar:2010we,Kostelecky:2016kfm}. While the details of these theories differ and those details matter for GW generation, they all predict modification to the propagation of GWs in a way that can be captured by simple parameterizations. By studying GW propagation, we can test new ideas about gravity without needing a complete theory. This allows us to set strict limits on parameters that tell us how much a theory might violate GR.

Here, we consider a test of GR performed by constraining a modified dispersion relation (MDR)~\cite{2012PhRvD..85b4041M}. In GR, GWs travel at the speed of light, resulting in a dispersion relation of the form $E^2=p^2c^2$. We can consider a generalized deviation from this relation, of the form
\begin{equation}
\label{eq:mdr_correction}
    E^2=(pc)^2+A_{\alpha}(pc)^{\alpha}\,,
\end{equation}
where $\alpha$ parametrizes the kind of deviation and the phenomenological amplitude parameter $A_\alpha$ quantifies the size of the deviation from GR. 
This type of dispersion relation is capable of capturing both low-energy ($\alpha<1$) and high-energy ($\alpha>1$) modifications to the dispersion, predicted by various alternative theories. 
For example, the case $\alpha =  0$, $A_0 > 0$ describes theories with massive graviton, such as \citet{2011PhRvL.106w1101D}. The LVK collaboration has performed this test on $\alpha \in \{0, 0.5, 1, 1.5, 2.5, 3, 3.5, 4\}$, with both positive and negative amplitudes, to capture a wide range of possible deviations (the case $\alpha = 2$ is excluded, as it is equivalent to a redefinition of the speed of gravity, so it does not result in dispersion).

Since the first implementation of the MDR test~\cite{2019PhRvD.100j4036A}, it has remained largely unchanged, with only the waveform approximant updated to more advanced versions between the observing runs. In this paper, we extend the test to cover more dispersion relations, describe in detail every improvement to the analysis process, rerun the updated test on all 43 events the test was run on in O1-O3b (using the publicly available strain data~\cite{2021SoftX..1300658A,2023ApJS..267...29A}), and compare our updated results with the results of the GWTC-3 testing GR paper~\cite{2021arXiv211206861T}.

After the initial analysis, two longstanding errors common in the field of GW data analysis were discovered: the use of incorrect calibration uncertainty (described in detail in ~\citet{Baka:2025bbb}) and the incorrect incorporation of data windowing to the likelihood function~\cite{2025arXiv250811091T}. All the results presented here were corrected for these errors in postprocessing by reweighting.

\subsection{Analysis overview}

The MDR analysis is based on Bayesian parameter inference. The posterior probability distribution of model parameters $\bm{\theta}$ conditioned on the observed data $\bm{d}$ is given by Bayes' law
\begin{equation}
    p(\bm{\theta}|\bm{d}) = \frac{\mathcal{L}(\bm{d}|\bm{\theta})\pi(\bm{\theta})}{\mathcal{Z}} \,,
\end{equation}
where $\mathcal{L}(\bm{d}|\bm{\theta})$ is the likelihood of the data realization given model parameters, $\pi(\bm{\theta})$ is the prior probability and $\mathcal{Z}$ is the normalization factor called the evidence
\begin{equation}
    \mathcal{Z} = \int\mathcal{L}(\bm{d}|\bm{\theta})\pi(\bm{\theta})d \bm{\theta} \,.
\end{equation}

For the purpose of this analysis, the data vector $\bm{d}$ consists of $N$ independent observations (1 for every GW event), which can be written as $\bm{d}=(\bm{d_1,d_2,}\cdots\bm{,d_N})$. Similarly, the model parameter vector can be decomposed as $\bm{\theta}=(\bm{\eta,\theta_1,\theta_2,}\cdots\bm{,\theta_N})$, where $\bm{\eta}$ is vector of model parameters shared between observations (GR violating parameters in case of MDR) and $\bm{\theta_i}$ is vector of parameters unique to event $i$. With this decomposition and using the independence of different observations, we get

\begin{align}
    p(\bm{\theta}|\bm{d}) &\propto \mathcal{L}(\bm{d_1,d_2,}\cdots\bm{,d_N}|\bm{\theta})\pi(\bm{\eta,\theta_1,\theta_2,}\cdots\bm{,\theta_N}) \nonumber\\
    &= \pi(\bm{\eta})\prod_{i=1}^N\mathcal{L}(\bm{d_i}|\bm{\eta,\theta_1,\theta_2,}\cdots\bm{,\theta_N})\pi(\bm{\theta_i}) \nonumber\\
    &= \pi(\bm{\eta})\prod_{i=1}^N\mathcal{L}(\bm{d_i}|\bm{\eta,\theta_i})\pi(\bm{\theta_i}) \nonumber\\
    &= \pi(\bm{\eta})^{1-N}\prod_{i=1}^N\mathcal{L}(\bm{d_i}|\bm{\eta,\theta_i})\pi(\bm{\eta})\pi(\bm{\theta_i}) \nonumber\\
    &\propto \pi(\bm{\eta})^{1-N}\prod_{i=1}^Np(\bm{\eta,\theta_i}|\bm{d_i}) \,. \label{eq:combining}
\end{align}

In the MDR analysis, we focus on the posteriors of GR violating parameters, $p(\bm{\eta}|\bm{d})$, which can be obtained by marginalization of the equation above:

\begin{align}
    p(\bm{\eta}|\bm{d}) &= \int d\bm{\theta_1}d\bm{\theta_2}\cdots d\bm{\theta_N}p(\bm{\theta}|\bm{d}) \nonumber \\
    &\propto \pi(\bm{\eta})^{1-N} \prod_{i=1}^N \int d\bm{\theta_i}p(\bm{\eta},\bm{\theta_i}|\bm{d_i}) \nonumber \\
    &\propto \pi(\bm{\eta})^{1-N} \prod_{i=1}^N p(\bm{\eta}|\bm{d_i}) \,,
\end{align}
i.e. the combined marginalized posterior is  proportional to marginalized posteriors of individual events, multiplied by $\pi(\bm{\eta})^{1-N}$ prior factor.

As is standard for parameter estimation (PE) of GWs, we assume Gaussian stationary noise, so the likelihood takes the form~\cite{2015PhRvD..91d2003V}
\begin{equation}
    \ln\mathcal{L}(\bm{d_i}|\bm{\eta},\bm{\theta_i}) = \langle\bm{d_i}-\bm{h}(\bm{\eta},\bm{\theta_i})|\bm{d_i}-\bm{h}(\bm{\eta},\bm{\theta_i})\rangle+\text{const} \,,
\end{equation}
where $\bm{h}(\bm{\eta},\bm{\theta_i})$ is a frequency-domain template waveform and $\langle \bm{a}|\bm{b}\rangle$ is the noise-weighted inner product defined as
\begin{equation}
    \langle \bm{a}|\bm{b}\rangle = 4\Re\left[ \int \frac{a(f)b^*(f)}{S(f)} df \right] \,,
\end{equation}
with $T$ the signal duration and $S(f)$ power spectral density (PSD) of the interferometer.

For the MDR test, we use a sampling algorithm to  estimate the individual event posteriors $p(\bm{\eta}|\bm{d_i})$ and then combine them to place bounds on MDR amplitudes $A_\alpha$. To better understand the difference between the analysis performed in the GWTC-3 testing GR paper~\cite{2021arXiv211206861T} and our analysis presented here, we compare the two analyses below. Each step is explained briefly, with the full explanations in Sec.~\ref{sec:improv}.

For the analysis performed in the GWTC-3 testing GR paper:
\begin{enumerate}
    \item Frequency domain \textsc{IMRPhenomXP}~\cite{2021PhRvD.103j4056P} waveform model is modified to include effects of MDR, derived by considering the propagation of different frequency components at particle velocity. The correction manifests itself as a frequency dependent phase term.
    \item For every event, 8 different MDR corrections are considered, for $\alpha \in \{ 0, 0.5, 1, 1.5, 2.5, 3, 3.5, 4\}$, i.e. corrections due to different $\alpha$ values are sampled discretely. For each case, PE is performed using \textsc{LALInference}~\cite{lalsuite} software.
    \item For every MDR exponent $\alpha$, two PE runs are launched---one for positive and one for negative phenomenological amplitude parameter $A_\alpha$. The sampling is performed in the parameter $\log_{10}\lambda_{\mathrm{eff}}$, with the corresponding uniform prior. Here $\lambda_{\mathrm{eff}}$ is the effective length scale corresponding to the amplitude $A_\alpha$~\eqref{eq:lambda}.
    \item The $\log_{10}\lambda_{\mathrm{eff}}$ samples are transformed to the corresponding $A_\alpha$ samples.
    \item The samples for negative and positive $A_\alpha$ are merged into a single set, and reweighted to account for different evidence $\mathcal{Z}$ of the positive and the negative branches.
    \item The samples are reweighted to correct for the difference between sampling prior flat in $\log_{10}\lambda_{\mathrm{eff}}$ and analysis prior flat in $A_\alpha$.
    \item Kernel density estimation is performed on the posterior samples to approximate $p(A_\alpha|\bm{d_i})$. A Gaussian kernel is assigned to every sample, which are then added together to produce a continuous distribution.
    \item Probability Distribution Functions (PDFs) from different events are combined together, to obtain the combined result $p(A_\alpha|\bm{d}) \propto \prod_i p(A_\alpha|\bm{d_i})$.
\end{enumerate}

With our improved and extended method, we take the following steps:

\begin{enumerate}
    \item Frequency domain \textsc{IMRPhenomXPHM}~\cite{2021PhRvD.103j4056P} waveform model is modified to include effects of MDR, derived by considering the propagation of different frequency components at \textit{group velocity} (Sec.~\ref{sec:group}).
    \item For every event, 10 different MDR corrections are considered, for $\alpha \in \{ -3, -2, -1, 0, 0.5, 1.5, 2.5, 3, 3.5, 4\}$---we extend the analysis to negative values of $\alpha$, associated to new physics at low energies/frequencies (Sec.~\ref{sec:negative}). For each case, PE is performed using the \textsc{Bilby}~\cite{2019ApJS..241...27A,2020MNRAS.499.3295R} software with the \textsc{Dynesty}~\cite{sergey_koposov_2023_8408702} sampler (Sec.~\ref{sec:Bilby}). 
    \item The sampling is performed in the parameter $A_{\mathrm{eff}}$, with a corresponding uniform prior. For the case $\alpha = 0$, an additional run sampling in $m_{g,eff}$ with a corresponding flat prior is performed (Sec.~\ref{sec:sampling}).
    \item The $A_{\mathrm{eff}}$ samples are transformed to the corresponding $A_\alpha$ samples.
    \item The samples are reweighted to a prior flat in $A_\alpha$.
    \item Kernel Density Estimation (KDE) is performed on the posterior samples to approximate the continuous PDF $p(A_\alpha|\bm{d_i})$.
    \item PDFs from different events are combined together, to obtain the combined result $p(A_\alpha|\bm{d}) \propto \prod_i p(A_\alpha|\bm{d_i})$.
\end{enumerate}

In Sec.~\ref{sec:improv}, we motivate our modifications, explaining why they should lead to an improvement over the GWTC-3 results. In Sec.~\ref{sec:results}, we test our analysis on injected signals and reanalyze GWTC-3 data. In Sec.~\ref{sec:conclusions}, we draw our conclusions and suggest future improvements.

\section{\label{sec:improv}Improvements}

\subsection{\label{sec:Bilby}\textsc{Bilby} software}
Up until GWTC-3, PE performed for the MDR analysis used the \textsc{LALInference} package~\cite{lalsuite,2019PhRvD.100j4036A,2021PhRvD.103l2002A,2021arXiv211206861T}. It is PE software programmed in C that was instrumental during the first two observing runs. Since then, it has been phased out in favor of the \textsc{Bilby}~\cite{2019ApJS..241...27A,2020MNRAS.499.3295R} package for LVK analysis. This Python-based package was developed with the goal of being more user-friendly, making it easier to learn and modify the software and the analysis. We take this opportunity to migrate the MDR codebase to a \textsc{Bilby}-compatible format.

We implement the MDR test as part of the Python-based \textsc{Bilby\_tgr}~\cite{colm_talbot_2024_10940210} package. It extends \textsc{Bilby} by providing code necessary to perform TIGER~\cite{2014PhRvD..89h2001A,2018PhRvD..97d4033M,Roy2026} (Test Infrastructure for GEneral Relativity), MDR, FTI~\cite{2023PhRvD.107d4020M} (Flexible Theory Independent Method), pSEOB~\cite{2018PhRvD..98h4038B,2021PhRvD.103l4041G,2023PhRvD.108b4043M,2025PhRvD.111l4040P} (\textsc{pSEOBNRv4HM} Ringdown Analysis), SSB~\cite{2021Univ....7..380O,2023PhRvD.107f4031H} (Spacetime Symmetry Breaking) and SIQM~\cite{2019PhRvD.100j4019K,2024PhRvD.109b3016D} (Spin-Induced Quadrupole Moment) analysis for testing GR. We kept the code structure as similar as possible to \textsc{Bilby}, making it easy to include the MDR effect on top of any current analysis. There are 3 important modules located in \verb|Bilby_tgr.mdr|:

\begin{itemize}
    \item \verb|source|---implements waveform source models that include the effect of MDR. Compatible with all frequency domain \textsc{LALSimulation} waveform approximants.
    \item \verb|conversion|---implements conversion functions between different parametrizations of MDR (like converting between the amplitude $A_0$ and the graviton mass $m_g$ parametrizations).
    \item \verb|postprocessing|---implements functions to reweight samples to different analysis priors.
\end{itemize}
Additionally, we made the MDR analysis compatible with the \textsc{Asimov}~\cite{Williams2023} package for automating GW parameter estimation by implementing it as the \verb|BilbyMDR| pipeline.

The \textsc{Bilby} package can perform PE with a variety of samplers. In this work, we have been exclusively using the \textsc{Dynesty} nested sampler~\cite{sergey_koposov_2023_8408702,2020MNRAS.493.3132S,2004AIPC..735..395S,10.1214/06-BA127,2016S&C....26..383B,2019PASP..131j8005B}. The accuracy of nested sampling algorithm is controlled by the number of live points---a set of samples replaced one by one during iterations of the algorithm. In our testing, we found 1200 live points to provide sufficient accuracy for the posteriors, but for safety, we chose 1500 live points when running the analysis on GWTC-3 events.

\subsection{\label{sec:group}Group velocity parametrization}
\subsubsection{Difference between particle velocity and group velocity parametrizations}
The MDR test, as implemented in the past GWTC papers, uses the correction derived in~\citet{2012PhRvD..85b4041M}. We summarize important steps here.

Starting with Eq.~\eqref{eq:mdr_correction}, we can rewrite it to obtain an equation for the particle velocity $v_p = c^2p/E$ of GWs~\cite{2012PhRvD..85b4041M}
\begin{equation}
\label{eq:phase_v}
    \frac{v_{p}}{c} = \sqrt{1 - A_\alpha E^{\alpha-2}} \approx 1 - \frac{1}{2}A_\alpha (hf)^{\alpha-2} \,,
\end{equation}
with $c$ the speed of light, $h$ the Planck constant, and $f$ the frequency of the signal. The equation is approximated to linear order by assuming small dispersion $A_\alpha (hf)^{\alpha-2}\ll 1$. 

Different frequency components travel at different speeds, resulting in the dispersion of the gravitational signal emitted by the source, ie. the waveform emitted at the source is not the same as the waveform arriving at the detector. This is intuitive since different frequency components arrive at the observer with different time separations compared to how they were emitted at the source. Compared to the GR case, the waveform accrues a frequency-dependent phase modification
\begin{equation}
    h(f) = h_{\mathrm{GR}}(f) e^{i \delta \Psi(f)} \,,
\end{equation}
where the phase correction $\delta \Psi(f)$ is given by~\cite{2012PhRvD..85b4041M}
\begin{equation}
\label{eq:phase_correction}
    \delta \Psi(f) = \begin{cases}\frac{\pi D_\alpha h^{\alpha-2}(1+z)^{\alpha-1}}{(\alpha-1)c}A_{\alpha}f^{\alpha-1} & \alpha  \neq 1\\\frac{\pi D_\alpha(1+z)^{\alpha-1}}{hc}A_{\alpha}\ln(\frac{\pi G \mathcal{M}f}{c^3})& \alpha = 1\end{cases} \,,
\end{equation}
where $z$ is the cosmological redshift and $\mathcal{M}$ is the chirp mass of the system. $D_\alpha$ is a distance measure defined as
\begin{equation}
    D_\alpha = (1+z)^{1-\alpha} \int^{t_a}_{t_e}a(t)^{1-\alpha}dt \,,
\end{equation}
where $z$ is the redshift, $a(t)$ is the cosmological scale factor,  $t_e$ the emission time of the signal, and $t_a$ the absorption time. It has a similar form to the luminosity distance $D_L$, and for a dark-energy-matter dominated universe they have the forms
\begin{equation}
        D_{\alpha} = \frac{c(1+z)^{(1-\alpha)}}{H_0}\int_0^z \frac{(1+\bar{z})^{(\alpha-2)}}{\sqrt{\Omega_M(1+\bar{z})^3+\Omega_\Lambda}}d\bar{z}
\end{equation}
\begin{equation}
        D_{L} = \frac{c(1+z)}{H_0}\int_0^z \frac{(1+\bar{z})}{\sqrt{\Omega_M(1+\bar{z})^3+\Omega_\Lambda}}d\bar{z} \,,
\end{equation}
with $\Omega_M$ the matter density parameter and $\Omega_\Lambda$ the vacuum energy density parameter.

While Eq.~\eqref{eq:phase_correction} is the most direct way of stating the result of the derivation in \citet{2012PhRvD..85b4041M}, it is not the form used in the MDR test performed in GWTC-3---there are two differences~\cite{2019PhRvD.100j4036A,2021PhRvD.103l2002A,2021arXiv211206861T}.

First, it is inconvenient to use $D_\alpha$ for PE. As sampling is usually performed in luminosity distance, obtaining $D_\alpha$ would require numerically solving the distance-redshift relation for redshift $z$ and then integrating to obtain $D_\alpha$. This is computationally expensive and would greatly slow down the sampling. Instead, an effective parameter $A_{\mathrm{eff}}$ was introduced
\begin{equation}
\label{eq:A_eff_def}
    A_{\mathrm{eff}} = \frac{D_{\alpha}}{D_L}(1+z)^{\alpha-1}A_{\alpha} \,,
\end{equation}
which modifies Eq.~\eqref{eq:phase_correction} to
\begin{equation}
\label{eq:mdr_phase_eff}
    \delta \Psi_{\alpha}(f)=\begin{cases}\frac{\pi D_L h^{\alpha-2}}{(\alpha-1)c}A_{\mathrm{eff}}f^{\alpha-1} & \alpha  \neq 1\\\frac{\pi D_L}{hc}A_{\mathrm{eff}}\ln(\frac{\pi G \mathcal{M}f}{c^3})& \alpha = 1\end{cases} \,.
\end{equation}

Second, the amplitude parameter $A_\alpha$ was instead expressed in terms of an effective wavelength parameter
\begin{equation}
\label{eq:lambda}
    \lambda_{\mathrm{eff}}=hc|A_{\mathrm{eff}}|^{1/(\alpha-2)} \,,
\end{equation}
leading to
\begin{equation}
\label{eq:old_param}
    \delta \Psi_{\alpha}(f)=\mathrm{sgn}(A_\alpha) \begin{cases}\frac{\pi D_L}{\alpha-1}\lambda_{\mathrm{eff}}^{\alpha-2}(\frac f c)^{\alpha-1} & \alpha  \neq 1\\\frac{\pi D_L}{\lambda_{\mathrm{eff}}}\ln(\frac{\pi G \mathcal{M}f}{c^3})& \alpha = 1\end{cases} \,.
\end{equation}

The above derivation rested on an important assumption---the graviton travels at particle velocity. This makes sense in light of the origin of the MDR test as an extension of massive graviton tests. GW propagation is interpreted as a stream of particles and for the massive graviton dispersion, group and particle velocities are equal, so the distinction is not important.

For a general MDR, this is no longer true. In \citet{2022JCAP...08..016E}, the authors have pointed out that the dispersion of GWs should be treated as a wave packet traveling with the \textit{group velocity} $v_g=d\omega/dk$. This gives a result consistent with solving the equation of the propagation of GWs using the Wentzel–Kramers–Brillouin (WKB) approximation---it corresponds to what is actually happening to the waveform. Using the modified dispersion equation~\eqref{eq:mdr_correction}, we get, up to linear order in $A_\alpha$
\begin{equation}
    \frac{v_{g}}{c} = 1 - \frac{1}{2}(1-\alpha)A_\alpha (hf)^{\alpha-2} \,.
\end{equation}

Compared with Eq.~\eqref{eq:phase_v}, the difference between the two is just a re-scaling of the phenomenological amplitude parameter $A_\alpha$ by a factor of $1-\alpha$, resulting in the same form of the MDR waveform as with particle velocity.

The re-scaling factor $1-\alpha$ singles out an exceptional case: $\alpha=1$. In that case, the group velocity reduces to the speed of light $c$ and there is no dispersion. The waveform is still modified though. The phase velocity does not vanish, so GWs accumulate a constant phase offset, independent of frequency. For a signal with only the dominant $(2,2)$-mode present, this corresponds to a redefinition of the phase at coalescence, and therefore cannot help to constrain the amplitude $A_1$. But if  higher modes (HMs) are present, this lifts the degeneracy, enabling us to test for constraints on $A_1$.

Taking the above into consideration, we follow \citet{2022JCAP...08..016E} and choose our new MDR phase modification to be
\begin{equation}
\label{eq:correction_group}
    \delta \Psi(f) = -\frac{\pi D_L h^{\alpha-2}}{c}A_{\mathrm{eff}}f^{\alpha-1} \,,
\end{equation}
where, similarly to Eq.~\eqref{eq:mdr_phase_eff}, we are using the effective amplitude $A_{\mathrm{eff}}$ to speed up sampling.

\subsubsection{\label{sec:alpha_1}More on the $\alpha=1$ case}
As MDR parameterized in terms of group velocity substantially differs from the parametrization in terms of partcle velocity only for $\alpha=1$, this case merits a more in-depth look.

The MDR waveform depends directly on the MDR phase offset $\delta\Psi$, which is a cyclic variable. If we consider data $\bm{d}$ from a single observation, the posterior $p$ conditioned on this data will necessarily reflect this symmetry: $p(\delta\Psi|\bm{d})=p(\delta\Psi+2\pi n | \bm{d})$, with $n$ an arbitrary integer.

To make statements about the physics of the possible dispersion, we should re-express the posterior in terms of the phenomenological amplitude parameter $A_1$, since the same value should be shared across multiple events, enabling us to combine constraints. By inverting Eq.~\eqref{eq:correction_group} for the case $\alpha=1$, and taking into account the cyclic nature of $\delta\Psi$, we get
\begin{equation}
    A_{1,n} = -\frac{hc\delta\Psi}{\pi D_\alpha} + \frac{2hc}{D_\alpha}n \,.
\end{equation}

The posterior on $A_1$ therefore splits into multiple branches $A_{1,n}$. Each branch carries equal probability mass, but different location (mean) and width (standard deviation).

Because the branches are spaced uniformly from $-\infty$ to $+\infty$ (with separation $2hc\left<1/D_\alpha\right>$) and each carries the same probability distribution, the resulting posterior on $A_1$ is not normalizable. We cannot therefore place a bound on possible deviation from GR---there would always be infinite probability mass outside of it.

The problem is worse still. Because of the uniform spacing of the branches, the posterior is approaching uniform distribution far away from $A_1 = 0$. As we combine multiple observations by multiplying their posteriors, we again end up with a combined uniform posterior (far away from $A_1=0$). The combined posterior is still not renormalizable. No matter how many events we combine, we still are not able to place a constraint on the $A_1$ amplitude parameter.

\begin{figure}
    \centering
    \includegraphics[width=0.9\columnwidth]{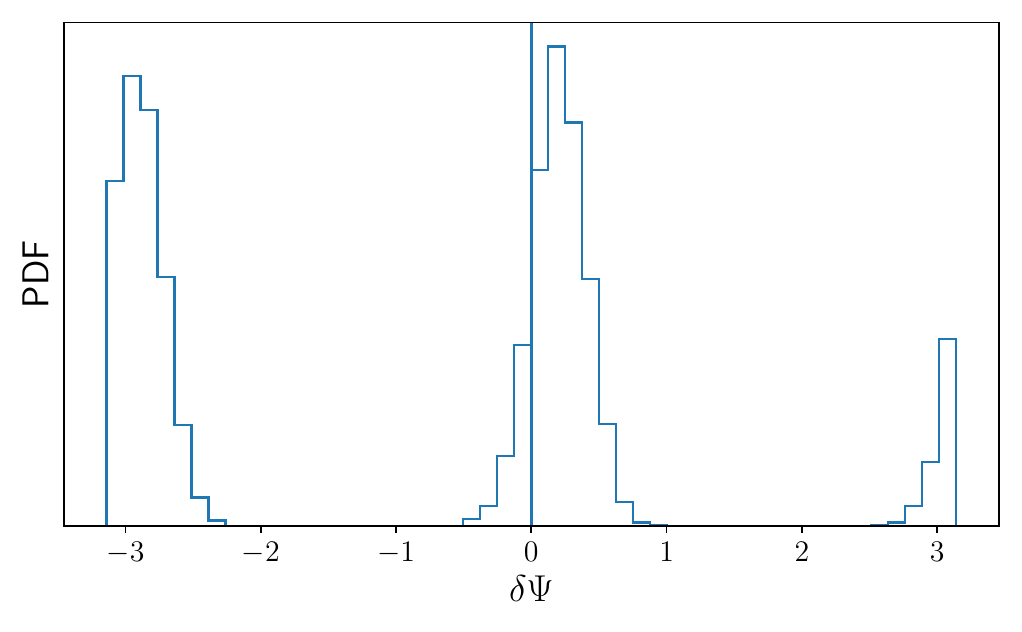}
    \caption{Posterior on the the MDR phase correction $\delta\Psi$ for a GW190412\_053044-like injection (GR). The HM content in the signal produces constraints on the possible values of $\delta\Psi$, in contrast to the uniform posterior expected from a signal with no HM content.}
    \label{fig:hm_phase}
\end{figure}

\begin{figure}
    \centering
    \includegraphics[width=0.9\columnwidth]{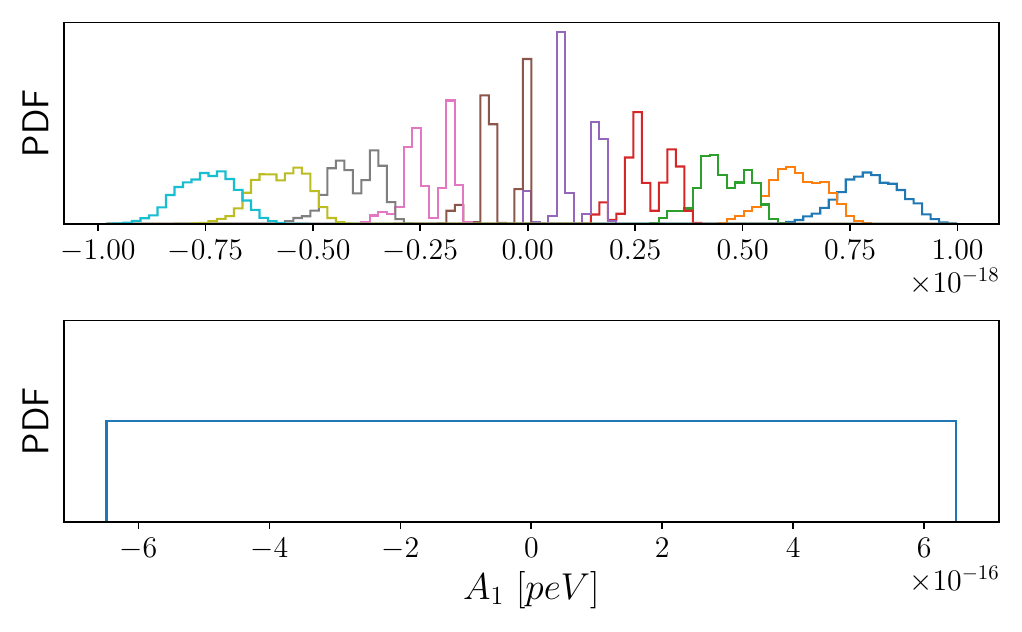}
    \caption{The posterior of the $A_1$ amplitude parameter for a GW190412\_053044-like injection. Top: Multiple values of $A_1$ correspond to the same correction to the waveform phase $\delta\Psi$, splitting the posterior between different branches. Bottom: On a large scale, this results in a uniform $A_1$ posterior.}
    \label{fig:branches}
\end{figure}

To illustrate the point, we injected a GW190412\_053044-like signal into the Livingston-Hanford-Virgo (LHV) detector network with Gaussian noise. We chose the injected MDR amplitude to be $A_1=0$. In Fig.~\ref{fig:hm_phase}, we can see the resulting posterior of $\delta\Psi$---the HM content is enough to constrain the parameter over uniform posterior expected with no HM present. In the upper panel of Fig.~\ref{fig:branches}, we plot the corresponding $A_1$ posterior, where for clarity we plotted only 5 branches, out of infinitely many. Near $A_1$ = 0, the peaks in the posterior are sharp and branches do not overlap. As we go further away, the peaks get broader, with the widths of different branches becoming more than the separation between them. While the peaks of different branches decrease away from $A_1=0$, suggesting that the posterior is concentrated near the origin, we have to keep in mind that the further away we move, the more branches overlap. Those two effects cancel each other exactly, which results in a uniform posterior (far away from $A_1$ = 0), shown in the bottom panel of Fig.~\ref{fig:branches}.

We reiterate that this problem with placing  bounds on the $A_1$ parameter is an inherent effect of the periodicity of the MDR correction for $\alpha=1$, and not an artifact of any sampling method---there is an infinite number of $A_1$ values that give the same phase correction. Combining multiple observations will not reduce bounds on $A_1$ (which cannot be placed in the first place), therefore we do not do so in this paper. This is in contrast to previous test of GR with MDR, where constraints on $A_1$ could be placed due to the parametrization by particle velocity.

\subsection{\label{sec:hm}Inclusion of higher order modes}
Until now, the MDR test was performed using waveform models containing only the dominant $(2,2)$-mode of the signal, with no HMs present (\textsc{PhenomPv2}~\cite{2019PhRvD.100b4059K} for GWTC-1~\cite{2019PhRvD.100j4036A} and \textsc{IMRPhenomXP}~\cite{2021PhRvD.103j4056P} for GWTC-2,3~\cite{2021PhRvD.103l2002A,2021arXiv211206861T}). This has not been a major problem, as of the 43 events the MDR test was run on, just two, GW190412\_053044 and GW190814\_211039, had significant subdominant modes present~\cite{2020PhRvD.102d3015A,2020ApJ...896L..44A}.

In this work, we reanalyze all 43 events the MDR test was run on in GWTC-1-3 with an updated waveform model, \textsc{IMRPhenomXPHM}~\cite{2021PhRvD.103j4056P}. For GW190412\_053044 and GW190814\_211039, we additionally analyze the events with \textsc{IMRPhenomXP} waveform, keeping all the other settings the same. This enables us to isolate the effect of the inclusion of HM on the posteriors from other improvements on the MDR test. We compare the posteriors in the results Sec.~\ref{sec:hm_results}.

\subsection{\label{sec:sampling}Sampling flat in $A_{\mathrm{eff}}$}

The MDR test performed in GWTC-3 was inefficient in the sampling of the parameter space. This can be traced to the MDR waveform being parametrized by Eq.~\eqref{eq:old_param} in terms of the effective wavelength parameter $\lambda_{\mathrm{eff}}$, with log-uniform sampling prior. This oversamples points near the GR value $A_\alpha = 0$, which then need to be rejected during the reweighting.

With our new method, the sampling is performed on the $A_{\mathrm{eff}}$ parameter, following Eq.~\eqref{eq:correction_group}. Compared to the old method, we do not have to split our PE between positive and negative $A_\alpha$. But it is the reweighting of the priors where we obtain the biggest improvement over the old method.

Let us consider the reweighting process in detail. We take a waveform parameterized during sampling by the parameters $\theta_i$, with the corresponding prior $\pi(\theta_i)$. The posterior samples are drawn from the posterior
\begin{equation}
    p(\theta_i|\bm{d})d\theta_i \propto \mathcal{L}(\bm{d}|\theta_i)\pi(\theta_i)d\theta_i \,,
\end{equation}
with $\mathcal{L}(d|\theta_i)$ the likelihood, $p(\theta_i|d)d\theta_i$ posterior probability mass corresponding to the sample and by $d\theta_i$ we mean $d\theta_1 d\theta_2 \cdots d\theta_N$. We are interested in the result parameterized by a different parametrization $\theta'_i$ under the corresponding prior $\pi'(\theta'_i)$. This means we have assigned the wrong probability mass to each sample during the sampling. We should have instead assigned
\begin{equation}
    p'(\theta'_i|\bm{d})d\theta'_i \propto \mathcal{L}(\bm{d}|\theta'_i)\pi'(\theta'_i)d\theta'_i \,.
\end{equation}
We can correct this discrepancy by assigning weights $w_i$ to the samples, representing the effective number of samples with the given parameters:
\begin{align}
    w_i &= \frac{p'(\theta'_i|\bm{d})d\theta'_i}{p(\theta_i|\bm{d})d\theta_i} = \frac{\mathcal{L}(\bm{d}|\theta'_i)\pi'(\theta'_i)d\theta'_i}{\mathcal{L}(\bm{d}|\theta_i)\pi(\theta_i)d\theta_i} \nonumber \\
    &= \frac{\pi'(\theta'_i)}{\pi(\theta_i)}\frac{d\theta'_i}{d\theta_i} = \frac{\pi'(\theta'_i)}{\pi(\theta_i)}\left|\frac{\partial\theta'_i}{\partial\theta_j}\right| \,,
\end{align}
where the last equality comes from the transformation law of volume elements (they transform by the determinant of the Jacobian matrix of the transformation).

For the MDR test performed in the GWTC-3, sampling was done in $\log_{10}\lambda_{\mathrm{eff}}$. This corresponds to vectors of parameters $\theta_i = (\log_{10}\lambda_{\mathrm{eff}}, D_L)$, $\theta'_i = (A_{\alpha}, D_L)$, where we omitted parameters unimportant for reweighting. The priors were chosen to be $\pi(\log_{10}\lambda_{\mathrm{eff}})=\pi'(A_{\alpha})=1$, with all the other parameters unchanged by the transformation, resulting in $\pi(\theta_i)=\pi'(\theta'_i)$. This led to the weights
\begin{align}
\label{eq:weights_old}
     w_i &= \left|\frac{\partial\theta'_i}{\partial\theta_j}\right| = 
     \begin{vmatrix}
\partial A_{\alpha} / \partial \log_{10}\lambda_{\mathrm{eff}}& \partial A_{\alpha}/ \partial D_L\\
\partial D_L / \partial \log_{10}\lambda_{\mathrm{eff}}& \partial D_L / \partial D_L
\end{vmatrix} \nonumber \\ &=
\begin{vmatrix}
\partial A_{\alpha} / \partial \log_{10}\lambda_{\mathrm{eff}} & \partial A_{\alpha}/ \partial D_L\\
 0 & 1
\end{vmatrix} = \left|\frac{\partial A_{\alpha}}{\partial \log_{10}\lambda_{\mathrm{eff}}}\right| \nonumber \\
&= \left|1/(\alpha-2)\frac{\partial \log_{10}A_\alpha}{\partial A_\alpha}\right|^{-1} \propto |A_\alpha| \,.
\end{align}

This poses a problem: for practical reasons, we cannot set the priors to extend over all the values of $A_\alpha$---we have to bound it by limiting values $A_{\mathrm{min}}$ and $A_{\mathrm{max}}$, or more to the point, $\log_{10}A_{\mathrm{min}}$ and $\log_{10}A_{\mathrm{max}}$. The upper bound poses no problem. As higher $A_\alpha$ corresponds to stronger deviation from GR, we expect the data to exclude the region above $A_{\mathrm{max}}$ from the posterior. But for $A_\alpha<A_{\mathrm{min}}$, the data offers no constraining power at all---all the values of $A_\alpha$ in the region are close enough to GR that the likelihood assigns the same probability to them. If we set the lower bound too low (compared with the width of the posterior), the reweighting will exclude most of the samples---the weights close to the lower bound $A_{\mathrm{min}}$, proportional to $|A_\alpha|$, will effectively be zero. This drastically reduces the effective sample size $n_{\mathrm{eff}}$. If we set the lower bound too high, we get railing in the posterior, manifesting after the reweighting as a dip in the posterior near $A_\alpha = 0$. Only $A_{\mathrm{min}}$ 2-3 orders of magnitude smaller than the width of the posterior would lead to a good posterior with large effective sample size and no artificial dips in PDF.

We found that this was the biggest issue with the MDR test done in GWTC-3 testing GR paper. Priors on $\log_{10}\lambda_{\mathrm{eff}}$ were too wide, resulting in low effective sample size $n_{\mathrm{eff}}$ for many events. This was especially pronounced for the $\alpha \in \{0, 0.5\}$ MDR tests, resulting in many multimodal posteriors.

Our new parametrization during sampling avoids this problem. With our choice of parameters and priors, we have $\theta_i = (A_{\mathrm{eff}}, D_L, ...)$, $\theta'_i = (A_{\alpha}, D_L, ...)$ and $\pi(\theta_i)=\pi'(\theta'_i)=1$. Following the same steps as in Eq.~\eqref{eq:weights_old}, we get

\begin{align}
     w_i & = \left|\frac{\partial A_{\alpha}}{\partial A_{\mathrm{eff}}}\right| \nonumber
= \left|\frac{\partial}{\partial A_{\alpha}}\frac{D_{\alpha}}{D_L}(1+z)^{\alpha-1}A_{\alpha}\right|^{-1} \\
&=\left|\frac{A_{\alpha}}{A_{\mathrm{eff}}}\right| \,.
\end{align}

For any given event, the uncertainty in the $D_L$ posterior is small on cosmological scales (ie. the range of redshift $z$ in the posterior samples is narrow). As both $D_L$ and $D_\alpha$ increase approximately linearly for small redshift, their ratio stays close to constant. This results in nearly equal weights, so we expect the effects of reweighting from $A_{\mathrm{eff}}$ to $A_\alpha$ to be minimal with our choice of sampling prior.

We can quantify the effect of reweighting on our samples by introducing sampling efficiency $\eta_{\mathrm{eff}}$ defined as
\begin{equation}
    \eta_{\mathrm{eff}} = \frac{\left(\sum_i^Nw_i\right)^2}{N\sum_i^Nw_i^2}
\end{equation}

where N is the total number of samples. We will use it later in Sec.~\ref{sec:reweighting_res} when assessing the impact of the reweighting process on the posteriors.

\subsection{\label{sec:graviton}Graviton mass posterior}

The MDR analysis tests different dispersion relations, distinguished by the value of the $\alpha$ parameter. The case $\alpha=0$ is of special interest as it corresponds to the massive graviton dispersion, where $A_0 = m_g^2c^4$ (this dispersion relation can also correspond to a charged graviton~\cite{2024JCAP...11..004N}). We can thus use the test to place constraints on the possible graviton mass $m_g$. The previous results~\cite{2021arXiv211206861T} suffer from the reweighting issue described in Sec.~\ref{sec:sampling}, so let us examine how to sample $m_g$ more efficiently.

One approach is to sample directly in $m_g$. The \textsc{Bilby} package, for which we implemented our test, can handle mismatches between different parametrizations of the prior and the waveform model, as long as an appropriate conversion function is provided. Similarly to sampling on $A_\alpha$, to avoid issues with having to calculate $D_\alpha(D_L)$ during sampling, we sample in the effective graviton mass parameter $m_{\mathrm{eff}}=\sqrt{(1+z)^{\alpha-1}D_{\alpha}/D_L}m_g$.

After the sampling, we follow the example of Eq.~\eqref{eq:weights_old} and reweight from prior $\pi(m_{\mathrm{eff}})=1$ to prior $\pi'(m_g)=1$, with weights:
\begin{align}
     w_i & = \left|\frac{\partial m_g}{\partial m_{\mathrm{eff}}}\right| \nonumber
= \left|\frac{\partial}{\partial m_{g}}\sqrt{\frac{D_{\alpha}}{D_L}(1+z)^{\alpha-1}}m_{g}\right|^{-1} \\
&=\frac{m_{g}}{m_{\mathrm{eff}}} \,,
\end{align}
where, similar to our sampling of the amplitude parameters, we expect this to lead only to negligible reweighting.

Sampling separately on $A_{0, eff}$ and $m_{\mathrm{eff}}$ is not necessary though. Since massive graviton dispersion is equivalent to dispersion with  $A_0>0$, we can obtain the posterior from our $A_{0, eff}$ samples. The weights needed to reweight from the sampling prior $\pi(A_{\mathrm{eff}})=1$ to the analysis prior $\pi'(m_g)=1$ are

\begin{align}
     w_i & = \left|\frac{\partial m_g}{\partial A_{\mathrm{eff}}}\right| \nonumber
= \left|\frac{\partial m_g}{\partial m_{\mathrm{eff}}}\frac{\partial m_{\mathrm{eff}}}{\partial A_{\mathrm{eff}}}\right| = \left|\frac{m_{g}}{m_{\mathrm{eff}}}\frac{\partial m_{\mathrm{eff}}}{\partial (m_{\mathrm{eff}}^2c^4)}\right| \\
& \propto \frac{m_{g}}{m_{\mathrm{eff}}} \frac{1}{m_{\mathrm{eff}}} = \frac{m_{g}}{m_{\mathrm{eff}}^2} \,.
\end{align}
These weights pose a problem due to a singularity at $m_g= m_{\mathrm{eff}} = 0$. We expect a finite posterior probability there (given consistency with GR), so samples from this region would get assigned infinite (or in practice very high) weights, which would result in a very low effective sample size $n_{\mathrm{eff}}$. Direct reweighting of samples is therefore not practical.

We are interested not in the samples though, but in the continuous PDF, which we need for combining posteriors from multiple observations. After obtaining the weights, we use them to perform a weighted Kernel Density Estimate (KDE) to get this PDF. We can solve the problem of singular weights by reversing the order of operations: we first perform KDE and then transform to the new prior.

To be specific, we start the process with the $p(A_0|\bm{d})$ distribution marginalized over all the other parameters $\bm{\theta}$, which is already smoothed out by the KDE and transformed to a prior flat in $A_0$. We wish to transform it to a posterior $p'(m_g|\bm{d})$ conditioned on the prior uniform in $m_g$. These posteriors can be written as
\begin{align}
    &p(A_0|\bm{d})dA_0 \propto \int d\bm{\theta} \mathcal{L}(A_0, \bm{\theta} |\bm{d})\pi(A_0)\pi(\bm{\theta})dA_0 \nonumber \,, \\
    &p'(m_g|\bm{d})dm_g \propto \int d\bm{\theta} \mathcal{L}(m_g, \bm{\theta} |\bm{d})\pi'(m_g)\pi(\bm{\theta})dm_g \nonumber \,, \\
    &p(A_0|\bm{d}) \propto p(A_0)\int d\bm{\theta} \mathcal{L}(A_0, \bm{\theta} |\bm{d})\pi(\bm{\theta}) \nonumber \,, \\
    &p'(m_g|\bm{d}) \propto p'(m_g)\int d\bm{\theta} \mathcal{L}(m_g(A_0), \bm{\theta} |\bm{d})\pi(\bm{\theta}) \nonumber \,, \\
    &\frac{p'(m_g|\bm{d})}{p(A_0|\bm{d})} \propto \frac{\pi'(m_g)}{\pi(A_0)} = 1 \nonumber \,, \\
    &p'(m_g|\bm{d}) \propto p(A_0(m_g)|\bm{d}) \,.
\end{align}
We see that the posterior does not transform under our change of variables, or to be more precise, the transformation of the posterior due to the change of variables cancels out with the simultaneous change of priors. In the results section (\ref{sec:individual_pos},~\ref{sec:combined}), we will compare the posteriors obtained by the two approaches outlined above.

\subsection{\label{sec:negative}Extension to negative $\alpha$s}
\subsubsection{Theoretical motivation}

When probing dispersion relations of the form~\eqref{eq:mdr_correction}, we ultimately test the speed of GWs as a function of frequency. We can make this more explicit in the following way:
\begin{align}
     E^2 &= p^2 c^2 + A_\alpha p^\alpha c^\alpha = p^2 c^2 \left[1 \pm \left(\frac{p}{p_\star}\right)^{\alpha - 2} \right], \nonumber \\ 
     &\Rightarrow c_{\rm gw}^2 = c^2 \left[1 \pm \left(\frac{f}{f_\star}\right)^{\alpha - 2} \right].
     \label{mdr_speed}
\end{align}
Here $p_\star$ and $f_\star$, implicitly defined in the above, correspond to momentum/frequency reference scales (here made explicit before being absorbed into the dimensionful  $A_\alpha$ parameter). Viewed in this way, it is clear that $\alpha > 2$ corresponds to high-frequency/high-energy modifications of the dispersion relation and hence the speed of GWs, while $\alpha < 2$ corresponds to a low-frequency modification. Models with a non-zero graviton mass ($\alpha = 0$) are an example of a well-known such low-energy (infrared) modification. 

\begin{figure}
    \centering
    \includegraphics[width=0.9\columnwidth]{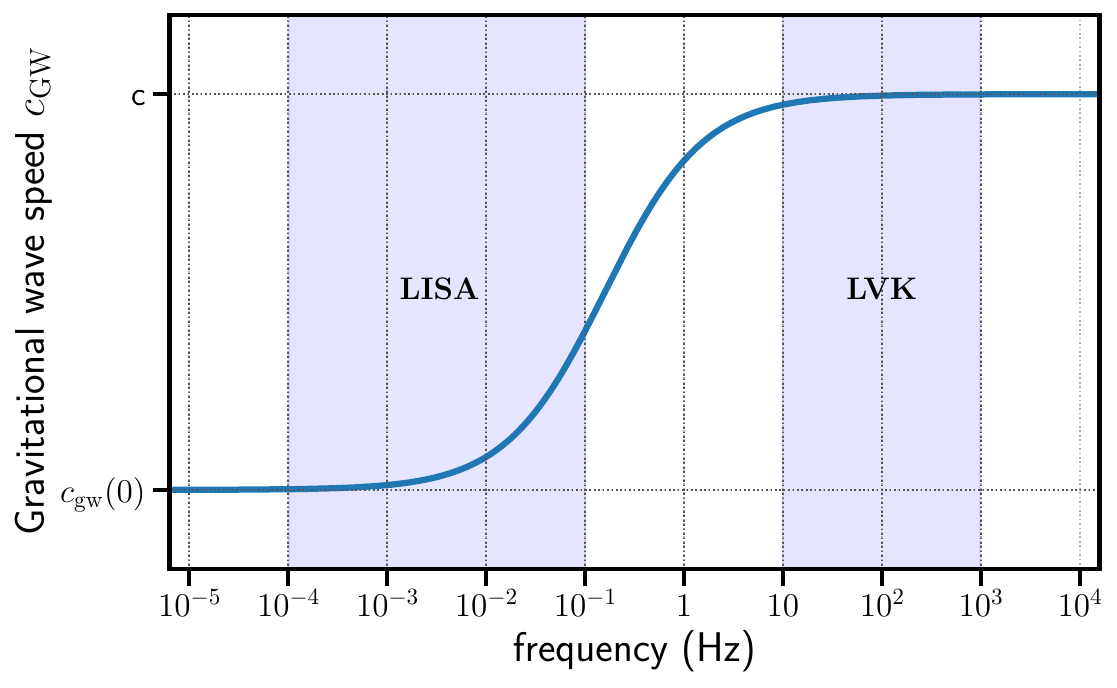}
    \caption{ Example frequency-dependence of the speed of GWs $\cgw$, where approximate LVK and LISA bands are highlighted.
    The behavior shown here is representative of what can be encountered in theories of dark energy that affect $\cgw$ on cosmological scales, with an asymptotic low-frequency speed $\cgw = \cgw(0)$ and a transition to $\cgw = c$ as one approaches the LVK band. The high-frequency tail of this transition is what the MDR analyses in the LVK band test and this corresponds to probing negative values of $\alpha$ in~\eqref{mdr_speed}. See the main text for further details.}
    \label{fig:cgw_freq}
\end{figure}

As discussed above, previous MDR analyses have focused on $\alpha \geq 0$ and we consider here in addition regions of parameter space with negative $\alpha$, i.e. additional low-frequency modifications. A key motivation for doing so comes from dark-energy related physics: if dark energy is dynamical in nature it is typically associated with one or more new particles/fields. Such fields would also couple to gravity and act as a `medium' that GWs propagate through. Naturally, this medium can affect the speed of those waves, and constraints from current cosmological data sets are consistent with ${\cal O}(1)$ deviations away from the speed of light~\cite{Bellini:2015xja,Noller:2018wyv}. However, dark energy theories that do affect $\cgw$ generically only describe frequency/energy scales up to at most ${\cal O}(10^2) \,\mathrm{Hz}$~\cite{deRham:2018red}.\footnote{This is analogous to how GR can only describe energy scales up to at most Planck energies, while a new high-energy completion is required to make predictions for higher energies. In the context of the dynamical dark energy theories discussed here, this `cutoff' is significantly lower than the Planck scale and can instead be located close to the energy scales probed by the LVK.} 

This means that, for such theories, an (unknown) high-energy completion of the fiducial new dark energy physics becomes relevant (and eventually dominates) as one approaches this cutoff, i.e. close to or somewhat below the LVK band.\footnote{Note that the cutoff is the largest possible energy/frequency scale, where the high-energy completion can take over, but this can already take place at significantly lower energies/frequencies. Theoretically predicting the precise scale would require detailed knowledge about such a fiducial (currently unknown) high-energy completion.}
As we know from existing LVK measurements, this high-energy completion will enforce $\cgw = c$ to high accuracy at frequency/energy scales in the LVK band and above. So, in the context of theories that significantly affect $\cgw$ on cosmological scales, one therefore naturally expects a frequency-dependent transition back to $\cgw = c$ upon approaching the LVK band (see Ref.~\cite{deRham:2018red,LISACosmologyWorkingGroup:2022wjo,Baker:2022eiz,Harry:2022zey}  for recent related studies of this frequency dependence). (The tail of) such a transition is what we can probe with dispersion relations, c.f. Fig.~\ref{fig:cgw_freq}. From the point of view of parametrizing dispersion relations in the LVK band, this can give rise to a negative $\alpha$ correction.

To see why, it is instructive to consider a template for such a transition~\cite{Harry:2022zey}
\begin{align}
\delta\cgw(f) = \delta\cgw^{(0)} \left(\tfrac{1}{2} - \tfrac{1}{2}\tanh\left[\sigma \cdot \log\left(f/f_\star\right)\right]\right).
\end{align}
Here $\delta\cgw(f) \equiv (\cgw(f) - c)/c$ denotes the fractional difference of the speed of GWs $\cgw$ from the speed of light $c$ as a function of frequency, $\delta\cgw^{(0)}$ is the low-frequency limit of this expression, $f_\star$ controls the frequency where the transition to $\cgw = c$ takes place, and $\sigma$ controls how quickly this transition takes place.  We will assume this expression to be small, i.e.  $\delta\cgw(f) \ll 1$.
Asymptotically, at frequencies larger than $f_\star$, one then finds
\begin{align}
f \gg f_\star:  \quad \delta \cgw \propto \left(\frac{f}{f_\star}\right)^{-2\sigma}  \quad \Rightarrow  \quad \alpha = 2-2 \sigma, 
\label{sTemp_powerlaw}
\end{align}
where we have indicated how $\sigma$ maps to $\alpha$ in this limit. If the transition is fast enough ($\sigma>1$, it will register as a dispersion with negative $\alpha$. Interestingly, sufficiently slow transitions with $\sigma = 1$ can therefore asymptotically mimic the $\alpha = 0$ case otherwise associated with the presence of a massive graviton. See \citet{Harry:2022zey} for further details on this and existing constraints on negative $\alpha$ using existing LVK (as well as forecasted LISA) observations. Note that the precise nature of this transition, and also the numerical value of the corresponding negative alpha coefficient, depends on the (unknown) nature of the high-energy completion and is, therefore, a currently unknown parameter in such theories to be measured.\footnote{As a related comment, note that while we test (half) integer values of $\alpha$ here, this should be interpreted as a coarse-grained search and not as motivated by underlying theoretical reasoning.} By probing dispersion relations with negative $\alpha$ in the LVK band, we are therefore testing and constraining the nature of dark energy in a complementary regime to that tested by cosmological observations. 

\subsubsection{Implementation details}

The model of low-energy modification discussed in the subsection above is compatible with the extension of our analysis to any $\alpha<0$, including arbitrary real values. For practical reasons, however, we must limit ourselves to a discrete set to probe\footnote{In principle, our implementation of the MDR analysis is compatible with setting a continuous prior on $\alpha$, but we have found this sampling method unstable in practice and were often unable to recover the injected parameters during sampling. This is likely due to the dimensions of $A_\alpha$ depending on $\alpha$.}. We have chosen to extend the MDR test to $\alpha \in \{ -3,-2,-1\}$, bringing the total number of tested dispersions to ten (We tried to keep the number low, as our analysis already requires significant computational resources for sampling.) Note that, contrary to cases of $\alpha>0$, we do not investigate half-integer values in the negative regime. This is because we found that our choice is a sufficient discretization, in the sense that we would pick up on possible deviations for any $\alpha \in [-3, 0]$; we show this in Sec.~\ref{sec:neg-alpha-results}.

Another problem that needed to be resolved was determining appropriate ranges for the uniform prior distributions used for $A_{\mathrm{eff}}$ and $A_{\alpha}$. Although we used GWTC-3 results to place appropriate prior bounds for $\alpha \geq 0$, we were unable to do that for negative $\alpha$ values. Therefore, in order to estimate the detectable order of magnitude of $A_{\alpha}$ for $\alpha<0$, we used the following argument.

If the true signal follows GR, then $A_{\alpha}$ must be small, in the sense that the MDR waveform is close to that which would be generated according to GR. The effect of dispersion is to accumulate additional phase $\delta\Psi_{\alpha}(f)$ compared to GR. If this phase is of the order of one full cycle, then the waveforms would be nothing alike and the posterior probability of $A_{\alpha}$ would go to 0. Therefore, the values for the posterior bounds on the amplitude parameter can be estimated using $\left|\delta\Psi_{\alpha}\left(f_{ref}, A_{\alpha}\right)\right| \sim 2\pi$, where $f_{ref}$ is some reference frequency dependent on the system parameters and detector PSDs. Substituting into Eq.~\eqref{eq:correction_group}, we get
\begin{equation} \label{eq:bound-scaling}
    \Delta A_{\mathrm{eff}}\left(\alpha\right) \sim \frac{2 \, h c}{D_{L}}\left(hf_{ref}\right)^{1-\alpha},
\end{equation}
where $\Delta A_{\mathrm{eff}}\left(\alpha\right)$ is the width of $p\left(A_{\mathrm{eff}}\left|\,\bm{d},\alpha\right.\right)$, the posterior of the effective amplitude parameter for a given $\alpha$. Comparing widths for different dispersions, we get a scaling relationship
\begin{equation} \label{eq:alpha_scaling}
    \frac{\Delta A_{\mathrm{eff}}\left(\alpha\right)}{\Delta A_{\mathrm{eff}}\left(\alpha^{\prime}\right)} = (hf_{ref})^{\alpha^{\prime}-\alpha},
\end{equation}
which we use in order to set the prior ranges for $\alpha < 0$ using our priors for the $\alpha=0$ and $\alpha=0.5$ dispersions. Note that this should hold true only for $\alpha<1$ which gives low-frequency corrections to the phasing. The $\alpha>1$ dispersions give a high-frequency correction, so the reference frequency used in the derivation is likely to be different. We verify this scaling relationship in Sec.~\ref{sec:neg-alpha-results} for our 43 GWTC-3 events.

\section{\label{sec:results}Results}
\subsection{\label{sec:injection_res}Injection tests}
In our improved MDR test, we have changed the nested sampler, switched the parameters we sample over, and expanded the analysis to cover more possible dispersion relations. Before reanalyzing the real events, we perform an injection campaign to test the validity of our analysis.

We have chosen to analyze 100 BBH injections. We sample the injection parameters from standard \textsc{Bilby} PE priors, with 2 exceptions: we limited our chirp mass $\mathcal{M}_c$ to a $10.2-50\,M_\odot$ range and luminosity distance $D_L$ to a $1-750 \, \mathrm{Mpc}$ range. The former was done for computational considerations, to limit our signals to an $8 \, \mathrm{s}$ window; the latter was done to ensure only sufficiently loud signals are analyzed. We provide a summary of the chosen priors in Tab.~\ref{tab:pp_priors} in App.~\ref{app:injections}.

The signals were injected into a 3 detector HLV network, into Gaussian noise based on Advanced LIGO/Virgo design sensitivity noise curves~\cite{2018LRR....21....3A}. The minimum frequency was set to $20 \, \mathrm{Hz}$, while the maximum was set to $2048 \, \mathrm{Hz}$, based on the sampling frequency of $4096 \mathrm{Hz}$. The time window was set to $8 \, \mathrm{s}$. For the nested sampler, we chose 1200 live points, to get well-behaved posteriors in a reasonable time frame. The sampling priors were chosen to be the same as the injection priors.

For every tested value of $\alpha$ (type of dispersion), 100 $A_{\mathrm{eff}}$ samples were drawn from a uniform distribution---one for every simulated BBH signal. The widths of those distributions were chosen based on the posteriors from the GWTC-3 MDR test---wide enough to include any recovered value of $A_{\mathrm{eff}}$ for any individual event. Again, these injection priors are summarized in Tab.~\ref{tab:pp_priors}.

\begin{figure}
    \centering
    \includegraphics[width=0.9\columnwidth]{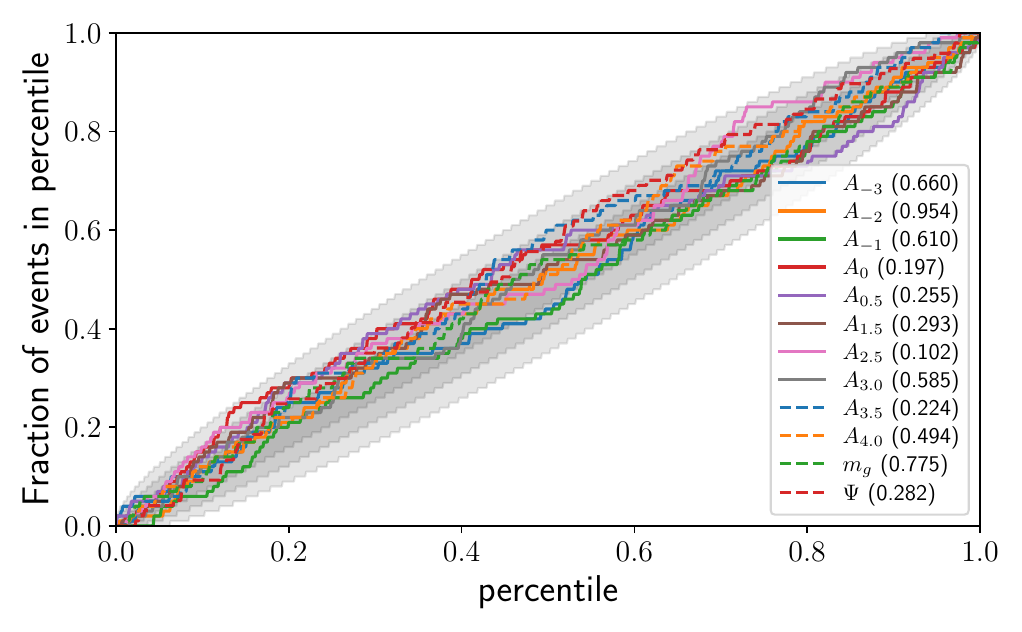}
    \caption{The pp-plot for the injection test of MDR. For clarity, we plotted just the sampled non-GR parameter ($A_{\mathrm{eff}}$, $m_{\mathrm{eff}}$ or $\delta\Psi_{MDR}$, depending on the injection set). The p-values were computed from KS statistics, and are consistent with a uniform distribution of percentiles. The combined p-value across the non-GR parameters is 0.49.}
    \label{fig:pp_combined}
\end{figure}

We found no problems with PE---we can always recover the injected $A_{\mathrm{eff}}$, as well as the GR parameters. We have additionally performed the percentile-percentile (pp) test, to investigate the self-consistency of our posteriors. The results are presented in Fig.~\ref{fig:pp_combined}. For clarity, we have chosen to show pp-plots only for $A_{\mathrm{eff}}$ parameters, with full results summarized in Tab.~\ref{tab:pp_values} in App.~\ref{app:injections}. The p-values for individual parameters were computed using the Kolmogorov–Smirnov test (KS test)~\cite{ks-test} and combined together using Fisher’s combined probability test~\cite{fisher1925statistical}.

We see that our posteriors are consistent with a uniform distribution of percentiles, with the lowest p-value 0.1 obtained for $\alpha=2.5$ and a combined p-value of 0.49. There is, in general, no problem with the recovery of GR parameters, with the notable exception of $\alpha=2.5$ injection set. This case has some of the lowest p-values, particularly for the time of coalescence $t_c$ (0.001) and primary spin magnitude $a_1$ (0.005). Refer to App.~\ref{app:injections} for a detailed discussion.

\subsection{\label{sec:hm_results}Higher order modes}

As mentioned in Sec.~\ref{sec:hm}, GWTC-3 has two events, GW190412\_053044 and GW190814\_211039, with significant HM contribution. As such, it is instructive to see how accounting for HM affects the posteriors for these events. We compare the posteriors obtained with \textsc{IMRPhenomXPHM} and \textsc{IMRPhenomXP} waveform models while keeping all the other sampler settings the same. We additionally compare it to the GWTC-3 results, rescaled to account for changing the particle velocity parametrization to group velocity.

\begin{figure}
    \centering
    \includegraphics[width=0.9\columnwidth]{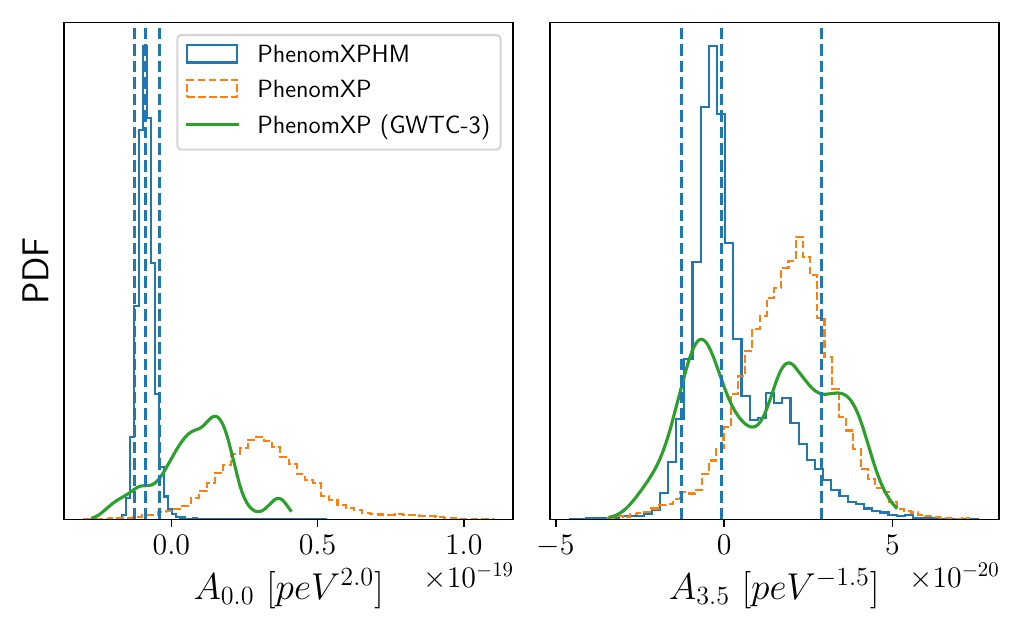}
    \caption{The GW190412\_053044 posteriors for $\alpha=0$ (left) and $\alpha=3.5$ (right) dispersions. Three posteriors are plotted: one obtained using the \textsc{IMRPhenomXPHM} waveform (blue), one using the \textsc{IMRPhenomXP} waveform (orange), and one from the GWTC-3 MDR analysis (green), also using the \textsc{IMRPhenomXP} waveform. Dashed vertical lines indicate the median and the 90\% CI. For $\alpha=0$, we see a drastic reduction in the width of the posterior, while the width stays similar for the $\alpha=3.5$ case. These are illustrative for other analyses of events with significant HM content.}
    \label{fig:hom_comparison}
\end{figure}

We show an illustrative result in Fig.~\ref{fig:hom_comparison}. Our posterior obtained with the \textsc{IMRPhenomXP} waveform is close to the GWTC-3 posterior, but without any secondary peaks, with similar width, and shifted in a few cases. Compared to our \textsc{IMRPhenomXP} posterior, our posterior obtained with \textsc{IMRPhenomXPHM} is narrower with the peak shifting closer to the GR value of $A_\alpha=0$. This behavior is expected, as we know the signal has an HM contribution. By neglecting it, we are intentionally mismodelling the signal. This is common for the two HM events across all the values of $\alpha$, but most pronounced for $\alpha < 1$. For GW190412\_053044, the $\alpha=0$ case posterior narrows enough that the GR value lies outside 90\% credible interval (CI). For $\alpha > 1$, both posteriors are more similar in shape, but still with noticeable narrowing when including HM.
Similar behavior occurs for the GR parameters of the signal (e.g. chirp mass $\mathcal{M}_c$).

For GW190412\_053044, we see a small multimodality in $A_\alpha$ posterior fo $\alpha = 3.5$ (and $\alpha=4.0$; not shown). This is unrelated to the presence of HMs. While $A_\alpha$ posterior is unimodal for most events, a few show bimodalities for either low or high values of $\alpha$. They persist across different sampler settings and random seeds and are accurate posteriors for the given realization of the interferometer data.

\subsection{\label{sec:reweighting_res}Effects of reweighting on the posteriors}

As we explained in Sec.~\ref{sec:sampling}, a poor choice of the sampling parameter in GWTC-3 resulted in a low effective sample size in the posteriors. For some events, as much as 95\% of the samples were lost during the reweighting process.

\begin{figure}
    \centering
    \includegraphics[width=0.9\columnwidth]{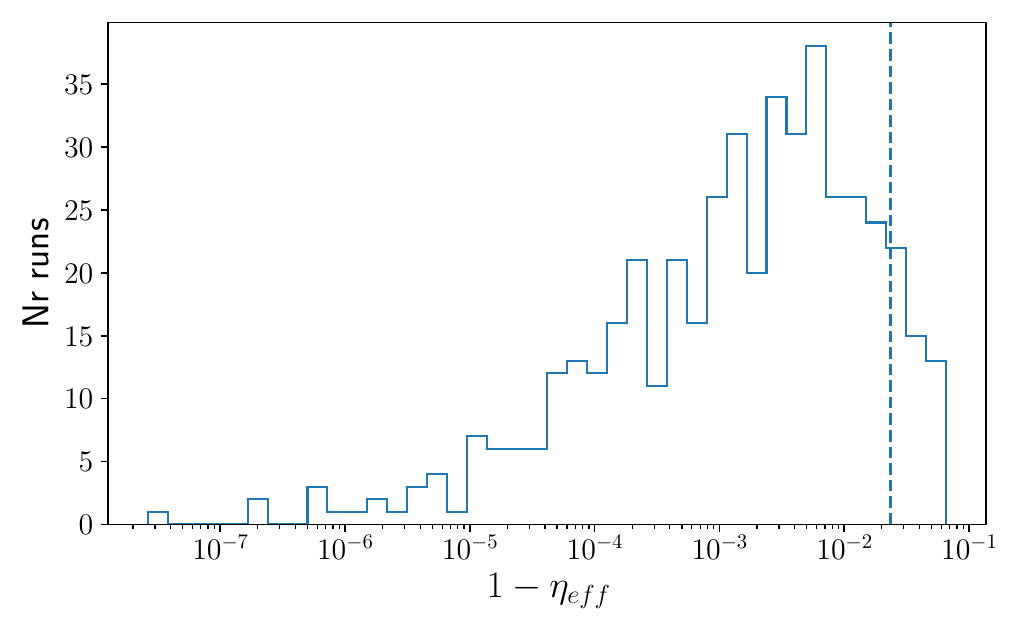}
    \caption{Resampling inefficiency (fraction of samples effectively rejected during resampling) $1-\eta_{\mathrm{eff}}$ when transforming from sampling posterior (uniform in $A_{\mathrm{eff}}$) to analysis posterior (uniform in $A_\alpha$). The dotted line indicates the 90\% percentile. The resampling is very efficient---in 90\% of the cases we keep more than 97.8\% of the samples.}
    \label{fig:sampling_eff}
\end{figure}

Our sampling in $A_{\mathrm{eff}}$ is much more efficient. The histogram in Fig.~\ref{fig:sampling_eff} describes the sampling inefficiency $1-\eta_{\mathrm{eff}}$, of our results---the effective fraction of rejected samples during reweighting process. Only in 10\% of the cases do we lose more than 2.2\% of the samples, with the worst-case scenario of losing 8\% of the samples. Compared to the issues encountered by the GWTC-3 analysis, this is minuscule and the effects of resampling on the posteriors are minimal.

\begin{figure}
    \centering
    \includegraphics[width=0.9\columnwidth]{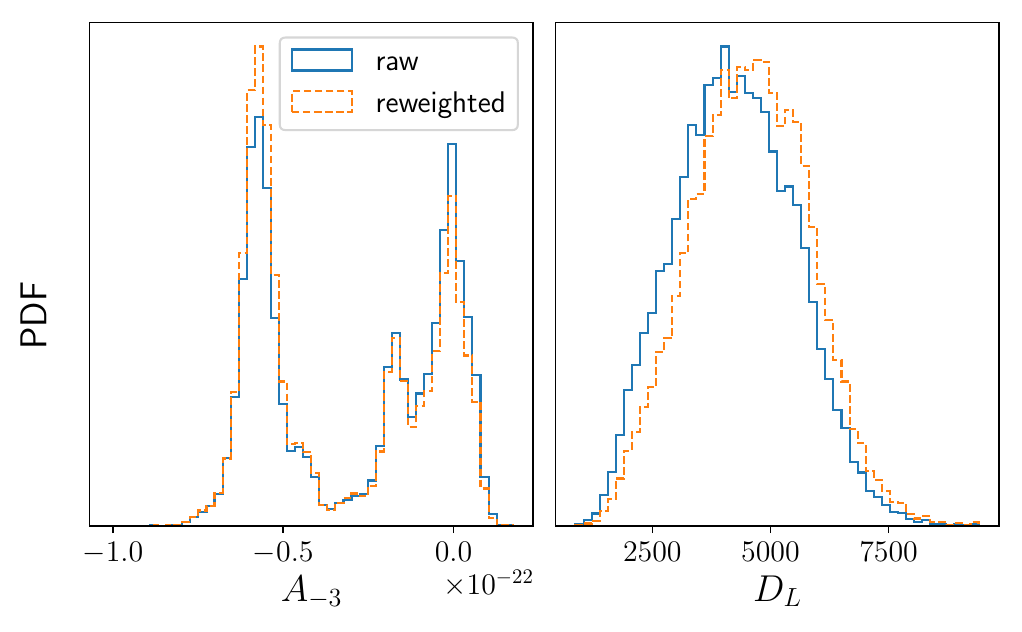}
    \caption{The posterior probability for the GW200219\_094415, $\alpha=-3$ PE run. This is the analysis with the worst resampling efficiency, $\eta_{\mathrm{eff}}=92\%$. Posteriors on $A_{-3}$ (left) and $D_L$ (right) are plotted, both before (blue) and after (orange) reweighting. Only a small change in posteriors is visible.}
    \label{fig:resampled_example}
\end{figure}

Consider the case GW200219\_094415,  $\alpha=-3$---the analysis with the worst resampling efficiency ($\eta_{\mathrm{eff}}=92\%$). In Fig.~\ref{fig:resampled_example}, we show two 1D posteriors for this event before and after resampling. The effect is small---in the $A_{-3}$ posterior there is a slight change in probability between the two peaks of the distribution. We stress that this is the worst-case scenario---for most of the reanalyzed events, there is no visible change between the posteriors before and after reweighting.

\subsection{Testing extension of MDR to negative $\alpha$\label{sec:neg-alpha-results}}
\subsubsection{Recovery with mismatched dispersion template}

As we already mentioned in Sec.~\ref{sec:negative}, we pick values of $\alpha\in\left\{-3, -2, -1\right\}$ as our extension of the MDR test to negative exponents. Since \textit{a priori} $\alpha$ could lie in between the chosen values, in this section we will investigate whether we can still recover a GR violation in this regime if we were to perform PE assuming a kind of dispersion (value of $\alpha$) different from the one we actually inject.

We performed three sets of injections into Gaussian noise with dispersions given by parameters $\alpha_{i}\in\left\{-0.5, -1.5, -2.5\right\}$, each consisting of the same 50 simulated BBH events. We chose the MDR amplitudes $A_{\alpha_i}$ such that each combined posterior shows a clear (but weak) deviation from GR.\footnote{$A_{\alpha_i}=0$ is recovered at the edge of the distribution---we aimed around the 0.01 quantile. If the injected deviation from GR was stronger, such that the GR value $A_\alpha=0$ lies completely outside the posterior distribution, then the small effect of mismodeling the dispersion would still show a GR violation.} Then, for each $\alpha_{i}$, we checked whether we again recover a deviation from GR if we assumed values of $\alpha_{r} \in \left\{\alpha_{i}\pm0.5\right\}$ (with consistent noise realization between each set of PE runs).

\begin{figure}
    \centering
    \includegraphics[width=0.9\columnwidth]{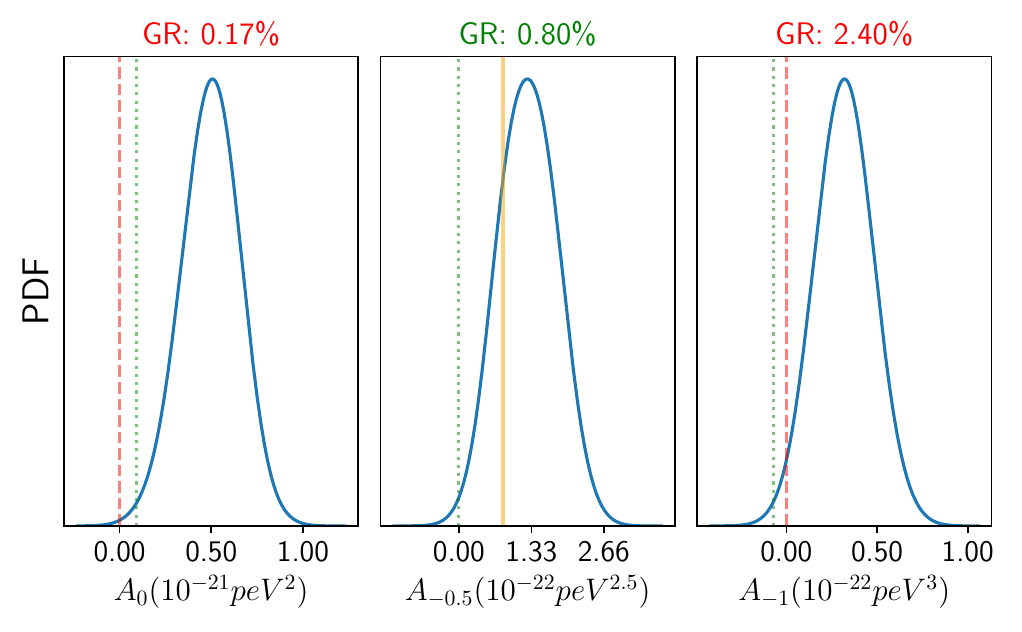}
    \caption{Combined posterior distributions of the MDR amplitude $A_{\alpha}$ for 50 simulated signals, with the injected dispersion given by parameters $\alpha=-0.5$ and $A_{-0.5}=8.1 \cdot 10^{-23}\, \text{peV}^{2.5}$ (orange, solid line). The injected GR violation is also recovered when assuming $\alpha=0$ (left) and $\alpha=-1$ (right), at percentiles $0.17\%$ and $2.40\%$ respectively (red, dashed lines). When instead we assume the injected $\alpha=-0.5$ (middle), it takes the value $0.80\%$ (green, dotted lines). Note the different energy dimensions for $A_{\alpha}$.}
    \label{fig:neg_alpha_pdfs}
\end{figure}

In Fig.~\ref{fig:neg_alpha_pdfs}, we show an example of this process for $\alpha_i=-0.5$ and $A_{-0.5}=8.1\cdot 10^{-23} \,\text{peV}^{2.5}$. The plots show combined $A_{\alpha}$ posteriors recovered assuming dispersions $\alpha \in \left\{0, -0.5, -1 \right\}$, respectively. When correctly modeling the injected dispersion ($\alpha=-0.5$), we recover the GR value of $A_\alpha$ at the $0.80\%$ percentile. If we perform PE of the same injection with a different dispersion model, we recover GR at the $0.17\%$ ($\alpha=0$) and the $2.40\%$ ($\alpha=-1$) percentile instead. The shift in percentiles is tiny, suggesting that mismodeled dispersion would still correctly recover a deviation from GR (for $\alpha=0$ the deviation is even slightly stronger than the true one).

We saw similar behavior for the other two injection sets as well: instead of the GR percentile of $0.59\%$ $\left(0.62\%\right)$ when recovering the parameters with the injected $\alpha_{i}=-1.5$ $\left(-2.5\right)$, when assuming dispersion with $\alpha_{r}=\alpha_{i}\pm0.5$, we recover GR at the $0.23\%$ and the $2.24\%$ (the $0.24\%$ and the $1.44\%$) percentile respectively---again, we pick up on the injected GR violation.

\begin{figure}
    \centering
    \includegraphics[width=0.9\columnwidth]{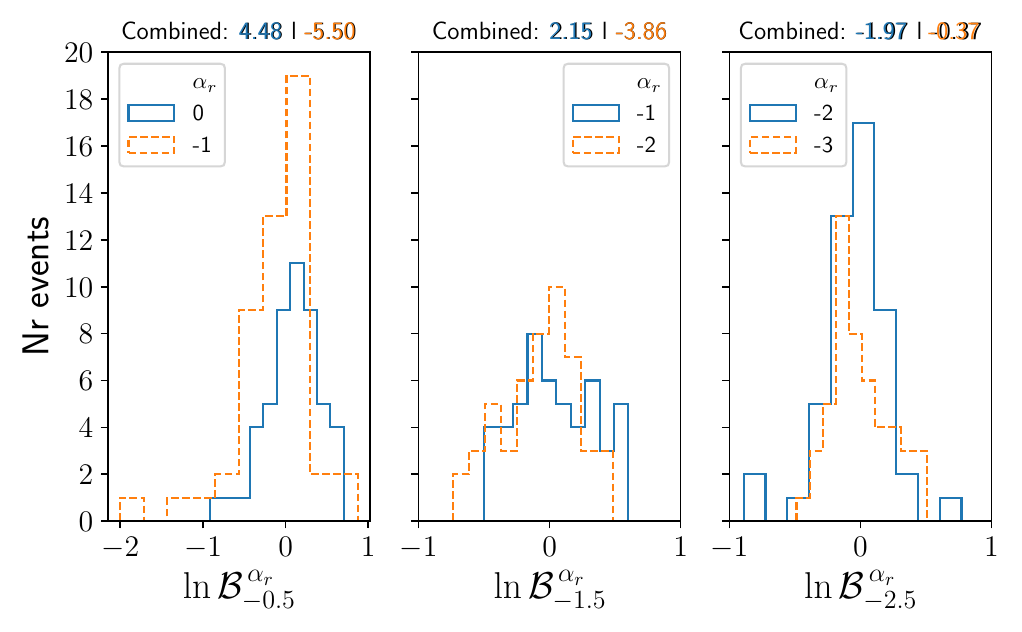}
    \caption{Histograms showing how the Bayes factors $\ln\mathcal{B}^{\alpha_{r}}_{\alpha_{i}}$ change when recovering the injected GR violation assuming different form of the dispersion. We compare mismatched form of the dispersion $\alpha_r\in\left\{ \alpha_i \pm 0.5 \right\}$ with the true injected dispersion $\alpha_i\in\left\{ -0.5, -1.5, -2.5 \right\}$. There is no strong preference between hypothesises for individual events. Above the histograms we show combined Bayes factors, each with uncertainty $\sigma=1.93$.}
    \label{fig:neg_alpha_deg_hist}
\end{figure}

Next, we compare the Bayes factors $\mathcal{B}^{\alpha_{r}}_{\alpha_{i}}$ between different hypotheses (different dispersion relations in recovery templates), with  $\alpha_i$ being the actual injected value and $\alpha_r$ the value used in the recovery. We can see the results in Fig.~\ref{fig:neg_alpha_deg_hist}, where we plot the individual as well as the combined Bayes factors for all 50 simulated events.

In each case, we see no significant preference when looking at individual events: values of $\ln\mathcal{B}^{\alpha_{r}}_{\alpha_{i}}$ lie in between $-1.37$ and $0.87$. When combining all 50 events together, for $\alpha_i\in\left\{-0.5, -1.5, -2.5\right\}$, the respective values of $\ln\mathcal{B}^{(\alpha_i+0.5)}_{\alpha_i}$ are $4.48$, $2.15$ and $-1.97$, while for $\ln\mathcal{B}^{(\alpha_i-0.5)}_{\alpha_i}$ they are $-5.50$, $-3.86$ and $-0.37$. To within the uncertainty on the combined Bayes factor, $\sigma=1.93$, there is no strong preference towards any value of $\alpha$. This is consistent with the combined $A_\alpha$ posteriors discussed above, where posteriors are shifted just slightly when recovering them with different $\alpha$.

All this leads us to conclude two things. Firstly, probing at $\alpha\in\left\{-3, -2, -1\right\}$ should be sufficient to detect a possible GR violation for any exponent between values of $-3$ and $0$; the GR value of $A_\alpha$ would still be recovered outside the bulk of the posterior. Secondly, in the case a deviation is picked up on, it may be hard to determine the specific nature of the dispersion. If, for example, we were to conclude that there is evidence for dispersion at $\alpha=-1$, we would have to perform further testing to rule out the possibility of $\alpha=-1$ masking as another dispersion. The corresponding constraints on the amplitude parameter $A_\alpha$ would, in contrast, strongly depend on the value of $\alpha$ we choose.

\subsubsection{Confirming hypothesis on $\Delta A_\alpha$ scaling}

In Sec.~\ref{sec:negative}, we have predicted that the width of the posterior distribution of the effective MDR amplitude parameter $\Delta A_{\mathrm{eff}}(\alpha)$ should scale as a power law with the exponent ``$-\alpha$'' for any given event. We confirm this hypothesis here using the posteriors of all 43 GWTC-3 events we analyzed in this paper.

Consider again Eq.~\eqref{eq:bound-scaling}. Taking the natural logarithm of both sides, we establish a linear dependency of $\ln\Delta A_{\mathrm{eff}}(\alpha)$ on $\alpha$, with the slope ``$-\ln \left(hf_{ref}\right)$'' (common for different MDR dispersions for any given event, but not necessarily between different events)
\begin{equation}
    \ln \Delta A_{\mathrm{eff}}\left(\alpha\right) \sim - \ln \left(hf_{ref}\right) \alpha + \text{const} \,.
\end{equation}

In practice, we define here $\Delta A_{\mathrm{eff}}(\alpha)$ as the width of the 90\% CI of the posterior $p\left(A_{\mathrm{eff}}\left|\,\bm{d},\alpha\right.\right)$. We then perform a linear regression of $\ln\Delta A_{\mathrm{eff}}(\alpha) \left[ \text{peV}^{2-\alpha} \right]$ as a function of $\alpha$  in the regime $\alpha<1$ to confirm the predicted scaling of the posterior widths.

\begin{figure}
    \centering
    \includegraphics[width=0.9\columnwidth]{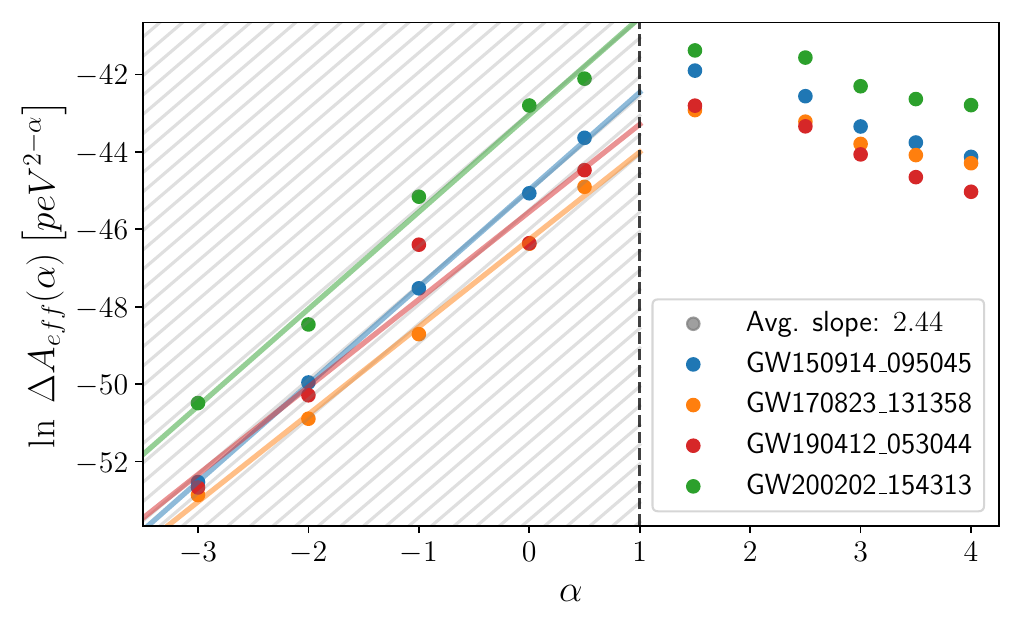}
    \caption{Scaling of the width of the $A_{\mathrm{eff}}$ posterior as a function of considered form of dispersion $\alpha$ for four illustrative events from GWTC-3. The logarithm of width scales linearly with $\alpha$, indicating a power law relationship. The light grey lines in the background correspond to the average slope between the events. The worst fit to a linear function among all 43 GWTC-3 events is GW190412\_053044 (shown above in red).}
    \label{fig:neg_alpha_slope_eg}
\end{figure}

We show some illustrative examples of the fits in Fig.~\ref{fig:neg_alpha_slope_eg}. We see that, for most events, a linear function is a good fit to our data, indicating that the widths of $A_{\mathrm{eff}}(\alpha)$ posteriors are indeed well described by the proposed power law. The event GW190412\_053044 is an exception and was included as it shows the worst fit to the power law among all 43 events. The slope factor ``$-\ln \left(hf_{ref}\right)$'' is in a similar range for all of the considered events, with the average value of $2.44$, minimum of $1.88$ and maximum of $2.72$. (The respective reference frequencies are $21 \, \mathrm{Hz}$, $37 \, \mathrm{Hz}$ and $16 \, \mathrm{Hz}$.) This agrees with our prediction that for each event the posterior widths scales as $\Delta A_{\mathrm{eff}}\left(\alpha\right) \sim \left(hf_{ref}\right)^{1-\alpha}$ with different $f_{ref}$ between events.\footnote{The reference frequencies are all around $20 \, \mathrm{Hz}$, which is the lower cutoff frequency in our analysis. MDR gives the largest modification to the phase of the waveform at this frequency, for $\alpha<1$.}

For our extension of the MDR test, we stopped at $\alpha=-3$. With the results above, we can extrapolate what constraint we could place on the $A_\alpha$ parameter if we were to extend our analysis to other negative powers.

\subsection{\label{sec:individual_pos}Posteriors of individual GWTC3-events}

In this section, we compare our new results with the results from GWTC-3 for all 43 events included in the MDR analysis. Before the comparison, we must stress two points, related to us changing the parametrization from particle velocity to group velocity.

First, as explained in Sec.~\ref{sec:group}, changing the parametrization rescales the results from GWTC-3: $A_\alpha \Rightarrow A_\alpha/(1-\alpha)$. This has a double effect of automatically narrowing all the $A_\alpha$ posteriors (apart from the $\alpha=1.5$ case where the scaling factor is $1/(1-1.5)=-2$) and reflecting them for $\alpha>1$. When comparing our result to the GWTC-3 results, we mean GWTC-3 results transformed this way, unless noted otherwise.

Second, we do not perform a comparison for the $\alpha=1$ dispersion. As demonstrated in Sec.~\ref{sec:group}, parametrization in terms of the group velocity means we can no longer place any bounds on $A_1$, so comparison with GWTC-3 results is not possible.

\begin{figure}
    \centering
    \includegraphics[width=0.9\columnwidth]{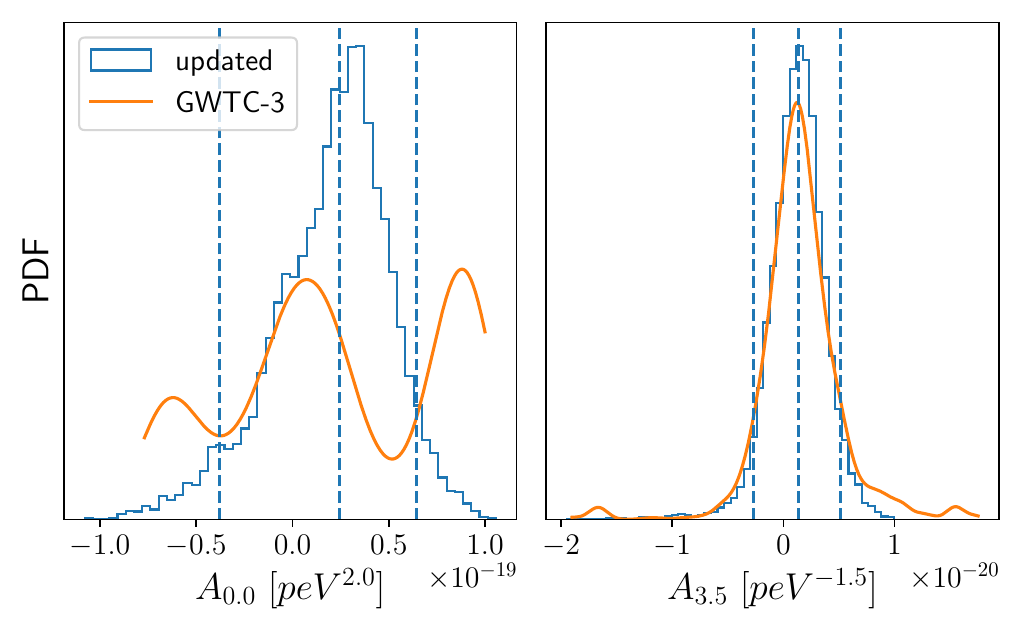}
    \caption{The $A_\alpha$ posteriors of GW191204\_171526. The new results (blue) are compared with GWTC-3 results (orange). The dashed lines indicate 0.05, 0.5, and 0.95 quantiles. Left: The low effective sample size in GWTC-3 often resulted in multimodal, sharply cutoff posteriors, corrected by our improved analysis, like for $\alpha=0$ dispersion. Right: In other cases, like for $\alpha=3.5$ our new results closely match the old ones. Both posteriors are consistent with GR.}
    \label{fig:gw191204}
\end{figure}

With the preliminaries out of the way, the changes in the posteriors between our results and GWTC-3 results fall into two broad categories---significant improvement or a minor difference. We will illustrate this on an example event, GW191204\_171526. The dispersion $\alpha=0$ gives a drastically better posterior than the GWTC-3 result, while $\alpha=3.5$ is consistent with the old result. Both can be seen in Fig.~\ref{fig:gw191204}.

For the $\alpha=0$ case, note the abrupt cut-off in the GWTC-3 posterior---it is not associated with any railing in the prior. The Kernel Density Estimate (KDE) performed for GWTC-3 results was bounded---PDF outside the range determined by the minimum and maximum sample was forced to be zero. If these samples come from the tail of the true underlying distribution, this does not cause any problems---probability density near these samples is close to zero, so bounding the KDE has almost no effect on the resulting distribution. But with the low effective sample size $n_{\mathrm{eff}}$ caused by the reweighting, the extremal samples can come from the region of high probability density. The bounded KDE introduces sharp artificial cutoffs in these cases.

While this can drastically alter the shape of individual posteriors, it has no effect on the combined posteriors. With 43 events combined, the PDF near the edges of the distribution will quickly go to 0 and the posteriors would look the same either for a bounded or unbounded KDE.

Next, note the triple peak in the GWTC-3 posterior. Again, this is caused by the low $n_{\mathrm{eff}}$---we do not have enough samples to adequately represent a wide single peak. The clusters of the few samples in this peak we do get show as multiple smaller peaks.

Both of the above problems disappear with our MDR test---we obtain a unimodal posterior without any boundary problems. We find this behavior common for the $\alpha=0$ case, with 30/43 events showing similar improvements from GWTC-3 results. The case $\alpha=0.5$ follows a similar pattern, with 25/43 posteriors showing significant improvements. For $\alpha \in \{1.5, 2.5, 3, 3.5, 4\}$ only 16/43 events show noticeable improvements in the posterior, usually on a lesser scale than the $\alpha=0$ and $\alpha=0.5$ cases (multiple peaks and sharp cutoffs are less common for GWTC-3 results in this group). This behavior is again consistent with the problem being caused by reweighting. We explained in Sec.~\ref{sec:sampling} that the prior width has a significant effect on the final result, and each $\alpha$ had it own choice of prior. In GWTC-3, priors for $\alpha > 1$ were better chosen than for $\alpha < 1$---the latter had too wide bounds, resulting in very significant sample loss during reweighting. Even for $\alpha>1$, the posteriors of individual events vary in width, so for some of the events, the chosen priors were too wide and we can see improvement in the posteriors with our method.

For the second category of improvements (little to no improvements in the posteriors), consider the $A_{3.5}$ posterior in Fig.~\ref{fig:gw191204}. Both GWTC-3 and our results have a strong single peak at the same location. The main differences are tiny secondary peaks near the tails of the GWTC-3 posterior that are not present in our result. The new posterior is slightly narrower than the old, which is a common behaviour in this group. We never observe a case where our new posterior is noticeably worse than the old one.

For the extension of the MDR test to $\alpha \in \{-1, -2, -3\}$, the GR value $A_\alpha=0$ is distributed uniformly among the quantiles of the posterior, with the KS test returning p-values $0.70, 0.40, 0.99$ respectively. The posteriors are mostly unimodal, with just 6/43 events having secondary peaks. For those, the GR value is recovered in one of the peaks.

\begin{figure}
    \centering
    \includegraphics[width=0.9\columnwidth]{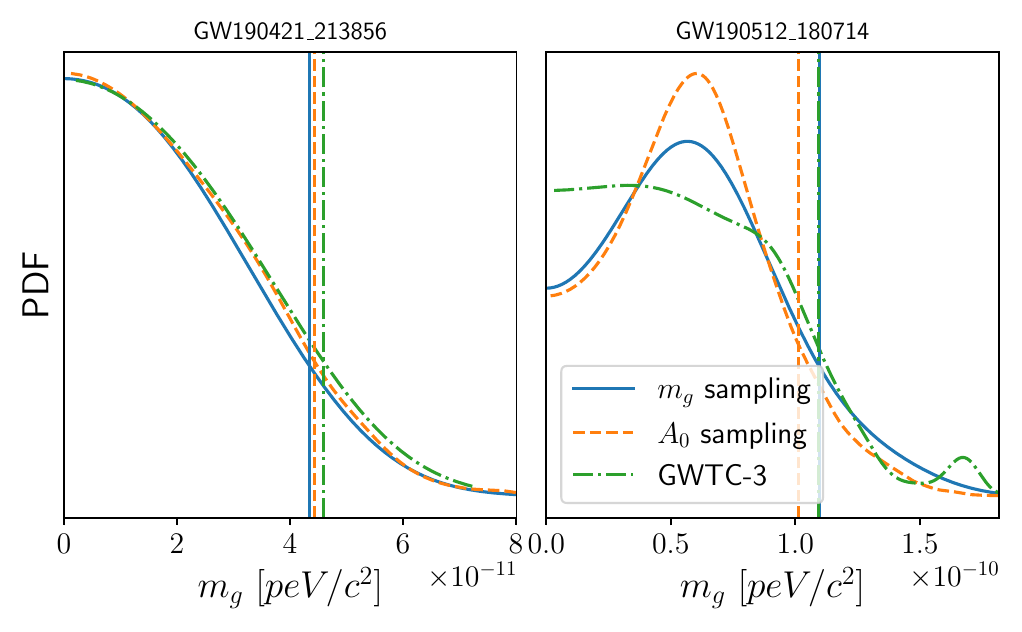}
    \caption{Comparison of graviton mass posteriors when sampling on $m_g$ (blue) or sampling on $A_0$ (orange) with the old GWTC-3 results (green). Left: For some events, like GW200129\_065458, all posteriors closely match each other. Right: For others, like GW190521\_074359,  the shapes of posteriors differ. Our new results are more similar to each other (between sampling methods) than to the old posteriors. Dashed lines indicate 90\% bounds on the graviton mass.}
    \label{fig:graviton_individual}
\end{figure}

As for the graviton mass posteriors, we mentioned in Sec.~\ref{sec:graviton} how we can obtain them by sampling either in $m_{\mathrm{eff}}$ or $A_{\mathrm{eff}}$ and then transforming the posterior. We compare the two approaches together with the GWTC-3 results. In Fig.~\ref{fig:graviton_individual} we show posteriors for GW190421\_213856, GW190512\_180714, which exhibit features common across events.

For 14/43 events, our new $m_g$ posteriors look very similar to the old GWTC-3 posteriors, with just minor changes in their shape (Fig.~\ref{fig:graviton_individual}, left). For the remaining 29 events, there is a clear difference between the old and the new results (Fig.~\ref{fig:graviton_individual}, right). This can again be attributed to the low effective sample size in GWTC-3.

The posteriors do not always peak at $m_g=0$ like for GW190421\_213856. In 16/43 cases, like for GW190512\_180714, posteriors peak away from 0. This is easily understood in terms of the $A_0 = m_g^2c^4$ parametrization. Assuming GR to be true, we expect the peak of the $A_0$ posterior to be centered equally likely on negative and positive values. When transforming to the $m_g$ posterior, negative values have to be rejected, so the peak moves to $m_g=0$, while in the positive case the peak survives and shows in the $m_g$ posterior away from 0.

Posteriors sampled on $m_{\mathrm{eff}}$ or $A_{\mathrm{eff}}$ follow the same shape, with posteriors matching almost exactly (like for GW190421\_213856) or showing noticeable deviation in some regions (like GW190512\_180714 near the peak of the posterior, Fig.~\ref{fig:graviton_individual}). In all cases, the posteriors are more similar to each other than to the equivalent GWTC-3 posterior. These discrepancies are most likely attributed to the finite sample size leading to uncertainty on the posteriors. In particular, transforming an $A_0$-sampled posterior to an $m_g$ posterior can drastically reduce the number of samples (we have to reject samples with $A_0<0$), making its shape more prone to errors.

\subsection{\label{sec:combined}Combined bounds on amplitude parameters and the graviton mass}
\begin{figure*}
    \centering
    \includegraphics[width=0.9\textwidth]{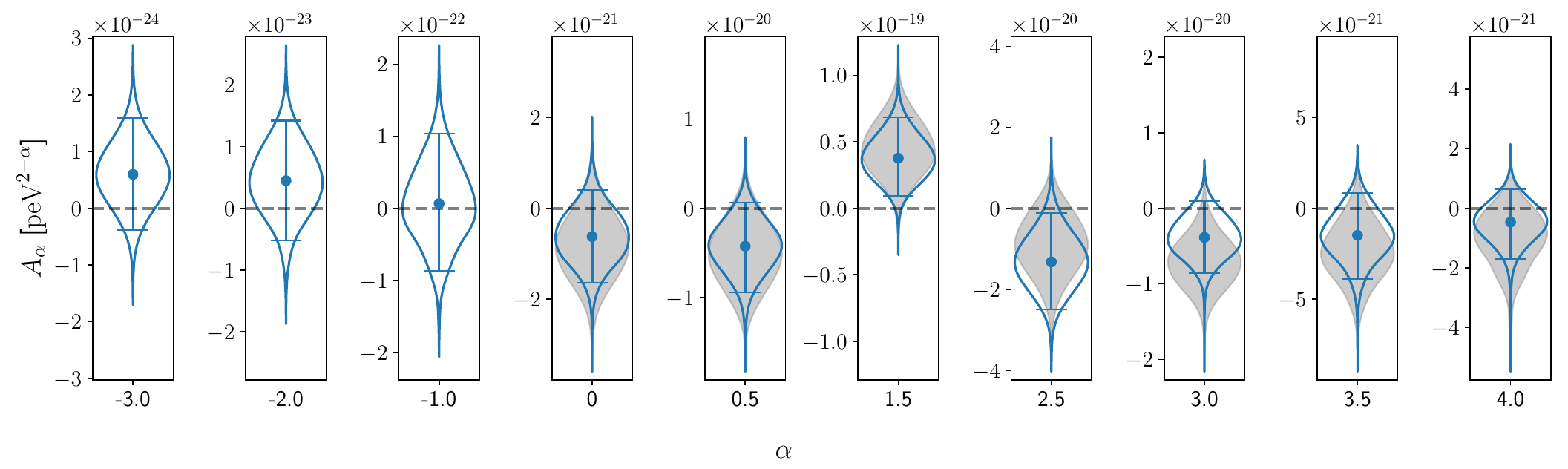}
    \caption{Posteriors on the amplitude parameters $A_\alpha$. Results from GWTC-3 (corrected for the change of parametrization to group velocity) are indicated by a shaded light-gray area, while our new results are represented by blue curves. The error bars indicate 90\% CIs. Posteriors are close to each other, but the new posteriors are on average 17\% narrower and shifted towards GR, apart from $\alpha=2.5$ dispersion which moves away from GR. For $\alpha=1.5$ and $\alpha=2.5$, GR lies outside the 90\% CI.}
    \label{fig:combined_amplitude}
\end{figure*}

For combining information from multiple events, we can use Eq.~\eqref{eq:combining} due to the independence of separate observations. As we have chosen a uniform prior, the combined posterior is just proportional to the product of the individual posteriors. We plot the result in Fig.~\ref{fig:combined_amplitude} and summarize the results in Tab.~\ref{tab:combined_amplitudes}.

We see that even with significant differences in individual posteriors between GWTC-3 and our results, the difference in combined posteriors is more modest. On average, we observe a 17\% reduction in the width of the combined $A_\alpha$ posteriors, with the smallest improvement of 6\% for $\alpha=2.5$ and the biggest improvement of 27\% for $\alpha=4$. This includes the effects of transforming GWTC-3 posteriors to the group velocity parametrization.\footnote{Without this process, our new posteriors for $\alpha=1.5$ and $\alpha=2.5$ are actually wider.} For $\alpha \in \{ 0.5, 1.5, 2.5 \}$, the quantile at which GR is recovered moves further away from the median in our results compared to GWTC-3 results. $\alpha = 1.5, 2.5$ are also the only cases where GR is recovered outside the 90\% CI, at quantiles 1.3\% and 96.3\% respectively.\footnote{This is not entirely unexpected---we are testing ten parameters, so just by statistical fluctuations we would expect around one parameter to be recovered outside $90\%$ CI.} Similarly to GWTC-3, where $\alpha = 1.5, 3$ recovered GR outside 90\% CI, this is primarily driven by GW200219\_094415 and GW200225\_060421, which recover GR near the edge of their posteriors. Without inclusion of these two events, all the combined posteriors would recover GR inside the 90\% CI. We did not find any problems in the analysis of these events and so kept them in the combined analysis. As the apparent tension with GR is driven by two extreme events, we expect it to disappear in the future, when more events are analyzed. For the newly tested cases of $\alpha \in \{-3,-2,-1\}$, GR is very well recovered with the median within $1\sigma$ of the GR value $A_\alpha=0$.

Notice that the worst deviations from GR (and the widest posteriors) occur near $\alpha = 1$ and $\alpha = 2$---dispersions we excluded as it is impossible to get any constraints on those parameters.

\begin{table*}[]
    \centering
    \begin{tabular}{ccc*{4}{cccc}ccc}
 \toprule
 & $m_g$  & & \multicolumn{3}{c}{$\bar{A}_{-3.0}$}  & & \multicolumn{3}{c}{$\bar{A}_{-2.0}$}  & & \multicolumn{3}{c}{$\bar{A}_{-1.0}$}  & & \multicolumn{3}{c}{$\bar{A}_{0.0}$}  & & \multicolumn{3}{c}{$\bar{A}_{0.5}$} \\
 \cline{4-6}
 \cline{8-10}
 \cline{12-14}
 \cline{16-18}
 \cline{20-22}
  & $90\%$ & & $5\%$ & $95\%$ & $Q_\text{GR}$ & & $5\%$ & $95\%$ & $Q_\text{GR}$ & & $5\%$ & $95\%$ & $Q_\text{GR}$ & & $5\%$ & $95\%$ & $Q_\text{GR}$ & & $5\%$ & $95\%$ & $Q_\text{GR}$ \\
 & $[10^{-11}]$  & & \multicolumn{2}{c}{$[10^{-24}]$} &  \% & & \multicolumn{2}{c}{$[10^{-23}]$} &  \% & & \multicolumn{2}{c}{$[10^{-22}]$} &  \% & & \multicolumn{2}{c}{$[10^{-21}]$} &  \% & & \multicolumn{2}{c}{$[10^{-20}]$} &  \%\\
 \midrule
GWTC-3 & 2.42 &  &  &  &  &  &  &  &  &  &  &  &  &  & -2.07 & 0.43 & 86.3 &  & -1.13 & 0.10 & 91.2 \\
updated & 2.21 &  & -0.38 & 1.59 & 15.7 &  & -0.53 & 1.42 & 22.3 &  & -0.86 & 1.04 & 45.6 &  & -1.63 & 0.41 & 84.1 &  & -0.94 & 0.07 & 92.2 \\
\toprule
 &   & & \multicolumn{3}{c}{$\bar{A}_{1.5}$}  & & \multicolumn{3}{c}{$\bar{A}_{2.5}$}  & & \multicolumn{3}{c}{$\bar{A}_{3.0}$}  & & \multicolumn{3}{c}{$\bar{A}_{3.5}$}  & & \multicolumn{3}{c}{$\bar{A}_{4.0}$} \\
 \cline{4-6}
 \cline{8-10}
 \cline{12-14}
 \cline{16-18}
 \cline{20-22}
 &   & & \multicolumn{2}{c}{$[10^{-20}]$} &  \% & & \multicolumn{2}{c}{$[10^{-20}]$} &  \% & & \multicolumn{2}{c}{$[10^{-20}]$} &  \% & & \multicolumn{2}{c}{$[10^{-21}]$} &  \% & & \multicolumn{2}{c}{$[10^{-21}]$} &  \%\\
 \midrule
GWTC-3 &  &  & 0.96 & 8.11 & 1.9 &  & -2.19 & 0.35 & 87.5 &  & -1.28 & -0.14 & 97.8 &  & -5.18 & 0.29 & 92.9 &  & -2.74 & 0.46 & 85.7 \\
updated &  &  & 0.95 & 6.83 & 1.3 &  & -2.50 & -0.11 & 96.3 &  & -0.85 & 0.09 & 90.4 &  & -3.89 & 0.85 & 85.0 &  & -1.69 & 0.64 & 75.3 \\
\bottomrule
\end{tabular}

    \caption{Combined results of the MDR analysis, compared with the GWTC-3 results. The table shows 90\% CIs for dimensionless graviton mass $\bar{m}_g = m_g/ (\text{peV}/c^2)$ and the dimensionless amplitude parameters $\bar{A}_\alpha = A_\alpha/\text{peV}^{2-\alpha}$. The quantiles of the GR hypothesis $Q_{\mathrm{GR}} = P(A_\alpha <0)$ are included as well.}
    \label{tab:combined_amplitudes}
\end{table*}

\begin{figure}
    \centering
    \includegraphics[width=0.9\columnwidth]{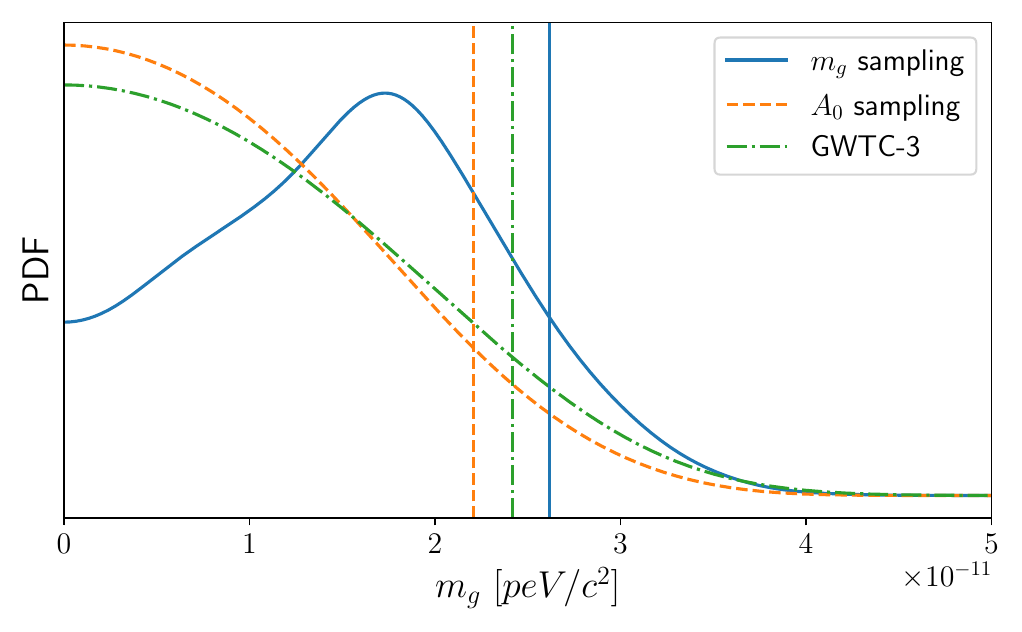}
    \caption{Combined graviton mass posteriors for the 43 GWTC-3 events. The posterior obtained from sampling in $m_g$ (blue) is compared with one obtained from sampling in $A_0$ (orange) and with the old GWTC-3 results (green). Dashed lines indicate 90\% bounds on graviton mass.}
    \label{fig:graviton_combined}
\end{figure}

In Fig.~\ref{fig:graviton_combined}, we show the combined graviton mass posterior, under a prior flat in graviton mass. We see that our new 90\% bound on the graviton mass moved from \num{2.42e-11} peV to \num{2.21e-11} peV (\num{2.62e-11} peV when sampling on $m_{\mathrm{eff}}$ instead of $A_{\mathrm{eff}}$). This relatively small change from the original results suggests that the effect of low effective sample sizes in individual GWTC-3 posteriors canceled out when combined, and the result is similar to the one obtained with our improved sampling method.

We note a concerning feature that the posterior has a peak away from zero when sampling in $m_{\mathrm{eff}}$ with a shape noticeably different than when sampling in $A_{\mathrm{eff}}$ and then transforming the samples. In principle, these posteriors should agree, and this is indeed the case on an event-by-event basis. However, KDE introduces different errors on the posteriors depending on in which space they are performed ($m_g$ vs. $A_0$). $m_g$ case has to contend with boundary effects near $m_g=0$ that distort the shape of the individual posteriors, which are then magnified when we combine the posteriors together. As a result, the posterior is less stable against the choice of the kernel bandwidth used than the posterior obtained from KDE in the $A_0$ space. As such, we decide to rely on the transformation of the $A_0$ posterior when quoting our combined bounds on graviton mass.

\section{\label{sec:conclusions}Conclusions}

We have improved on the MDR test performed in GWTC-3~\cite{2021arXiv211206861T} and verified on an injection campaign that it produces consistent, well-behaved posteriors. For many events, our method performs significantly better than the method employed in GWTC-3, giving more accurate posteriors with better effective sample sizes. This does not extend to combined results, with only a small difference between our result and GWTC-3. This reassures us that conclusions drawn from GWTC-3 about modified dispersion should remain correct.

For follow-up work, we plan to apply our MDR analysis in the tests of GR with the upcoming GWTC-4. With the amount of analyzed events expected to roughly double, we expect tighter constraints on the amplitude parameters and graviton mass. For dispersions showing currently a tension with GR values (like $\alpha=1.5$ recovering the GR value at the 1.8\% percentile), analyzing additional events might help to resolve it.

Our improvements to the MDR test might affect the MDR analysis with population-informed inference~\cite{2023PhRvD.108l4060P}. The authors of the paper show substantial improvement on the bounds of the graviton mass by incorporating population information about the distribution of astrophysical masses. Our results should more accurately recover the component masses (due to the low effective sampling size of GWTC-3 results), which can affect selection effects due to population models. Additionally, the authors had to combine the posteriors on $m_g$ in the log space for well-behaved results. We predict that with our results it should be possible to combine posteriors directly in $m_g$ space.

We have not done any testing of our improved method on binary neutron stars or neutron-star-black-hole systems. Those kinds of signals were excluded from GWTC-3 analysis due to computational constraints (long duration of the signals), and we ran into the same problem with parameter estimation taking too long. As these kinds of events are rarer and tend to be closer to us, excluding them should not have a significant effect on the combined posteriors. Nevertheless, in the future, we will investigate possible speed-up algorithms for the MDR test to make the analysis of long signals feasible.

\begin{acknowledgments}

We thank Manuel Piarulli and Sylvain Marsat for reviewing our BBH posteriors. We thank Chris Van Den Broeck and Elise S{\"a}nger for useful comments on the manuscript. The authors are grateful for computational resources provided by the LIGO Laboratory  and supported by National Science Foundation Grants PHY-0757058 and PHY-0823459. Plots were prepared with Matplotlib~\cite{Hunter:2007}. NumPy~\cite{harris2020array} and SciPy~\cite{2020SciPy-NMeth} were used during data analysis. TB is supported by the research program of the Netherlands Organisation for Scientific Research (NWO). JN is supported by an STFC Ernest Rutherford Fellowship (ST/S004572/1).

This research has made use of data or software obtained from the Gravitational Wave Open Science Center\footnote{\url{gwosc.org}}, a service of the LIGO Scientific Collaboration, the Virgo Collaboration, and KAGRA. This material is based upon work supported by NSF's LIGO Laboratory which is a major facility fully funded by the National Science Foundation, as well as the Science and Technology Facilities Council (STFC) of the United Kingdom, the Max-Planck-Society (MPS), and the State of Niedersachsen/Germany for support of the construction of Advanced LIGO and construction and operation of the GEO600 detector. Additional support for Advanced LIGO was provided by the Australian Research Council. Virgo is funded, through the European Gravitational Observatory (EGO), by the French Centre National de Recherche Scientifique (CNRS), the Italian Istituto Nazionale di Fisica Nucleare (INFN) and the Dutch Nikhef, with contributions by institutions from Belgium, Germany, Greece, Hungary, Ireland, Japan, Monaco, Poland, Portugal, Spain. KAGRA is supported by Ministry of Education, Culture, Sports, Science and Technology (MEXT), Japan Society for the Promotion of Science (JSPS) in Japan; National Research Foundation (NRF) and Ministry of Science and ICT (MSIT) in Korea; Academia Sinica (AS) and National Science and Technology Council (NSTC) in Taiwan.
\end{acknowledgments}

\section{Data availability}
All analyzed strain data are available from the Gravitational Wave Open Science Center~\cite{2021SoftX..1300658A,2023ApJS..267...29A}. The full results of the analysis, including the posterior samples, are available from \citet{baka_2026_18789951}.

\FloatBarrier

\appendix

\section{\label{app:injections}Full results of pp-test}

In Sec.~\ref{sec:injection_res}, we performed an injection campaign to test the self-consistency of posteriors obtained with our MDR test. In Tab.~\ref{tab:pp_priors} we provide the priors we used for drawing our injection parameters and used during sampling. The first part of the table details priors common for each tested modified dispersion, while the second part details priors on the non-GR parameters, unique for every tested dispersion.

\addtolength{\tabcolsep}{+2pt} 
\begin{table*}[]
    \centering
    \begin{tabular}{llrrl}
 \toprule
 parameter & prior & minimum & maximum & unit \\
 \midrule
$\mathcal{M}_c$ & Uniform in components & 10.2 & 50 & $M_\odot$ \\
$q$ & Uniform in components & 0.05 & 1 & - \\
$a_1, a_2$ & Uniform & 0 & 0.99 & - \\
$\theta_1, \theta_2$ & Sine & 0 & $\pi$ & - \\
$\phi_{12}, \phi_{JL}$ & Uniform & 0 & $2\pi$ & - \\
$D_L$ & Uniform source frame & 1 & 750 & Mpc \\
$\mathrm{DEC}$ & Cosine & $-\pi/2$ & $\pi/2$ & - \\
$\mathrm{RA}$ & Uniform & 0 & $2\pi$ & - \\
$\theta_{JN}$ & Sine & 0 & $\pi$ & - \\
$\psi$ & Uniform & 0 & $\pi$ & - \\
$\phi$ & Uniform & 0 & $2\pi$ & - \\
$t_c$ & Uniform & -0.1 & 0.1 & s \\
\midrule
$A_{-3}^{eff}$ & Uniform & $-1.7 \times 10^{-22}$ & $1.7 \times 10^{-22}$ & $\mathrm{peV}^{5}$ \\
$A_{-2}^{eff}$ & Uniform & $-2 \times 10^{-21}$ & $2 \times 10^{-21}$ & $\mathrm{peV}^{4}$ \\
$A_{-1}^{eff}$ & Uniform & $-2.5 \times 10^{-20}$ & $2.5 \times 10^{-20}$ & $\mathrm{peV}^{3}$ \\
$A_{0}^{eff}$ & Uniform & $-3 \times 10^{-19}$ & $3 \times 10^{-19}$ & $\mathrm{peV}^{2}$ \\
$A_{0.5}^{eff}$ & Uniform & $-2 \times 10^{-18}$ & $2 \times 10^{-18}$ & $\mathrm{peV}^{1.5}$ \\
$A_{1.5}^{eff}$ & Uniform & $-6 \times 10^{-18}$ & $6 \times 10^{-18}$ & $\mathrm{peV}^{0.5}$ \\
$A_{2.5}^{eff}$ & Uniform & $-2 \times 10^{-18}$ & $2 \times 10^{-18}$ & $\mathrm{peV}^{-0.5}$ \\
$A_{3.0}^{eff}$ & Uniform & $-9 \times 10^{-19}$ & $9 \times 10^{-19}$ & $\mathrm{peV}^{-1.0}$ \\
$A_{3.5}^{eff}$ & Uniform & $-7 \times 10^{-19}$ & $7 \times 10^{-19}$ & $\mathrm{peV}^{-1.5}$ \\
$A_{4.0}^{eff}$ & Uniform & $-1 \times 10^{-18}$ & $1 \times 10^{-18}$ & $\mathrm{peV}^{-2.0}$ \\
$m_g^{eff}$ & Uniform & 0 & $5.5 \times 10^{-10}$ & peV \\
$\Psi_{MDR}$ & Uniform & 0 & $2\pi$ & - \\
\bottomrule
\end{tabular}
    \caption{Priors chosen for our injection campaign and for generating pp-plots. $\mathcal{M}_c$ stands for chirp mass, $q$ for mass ratio, $a_i$ for the dimensionless spin magnitude of the binary component, with $\theta_i$ the corresponding angle between the spin and the orbital angular momentum; $\phi_{12}$ is the azimuthal angle separating the spin vectors and $\phi_{JL}$ is the cone of precession around the total angular momentum; $D_L$ is the luminosity distance, while DEC and RA are the declination and the right ascension; $\theta_{JN}$ is the angle between the binary's angular momentum and the line of sight, $\psi$ is the polarization angle and $\phi$ and $t_c$ are the phase and time at coalescence. Other parameters parametrize deviation from GR for different dispersions.}
    \label{tab:pp_priors}
\end{table*}
\addtolength{\tabcolsep}{-2pt}

In the main body of the paper, we provided results of the  pp-test only for GR-violating parameters, as those are of interest for performing MDR tests. In Tab.~\ref{tab:pp_values} we present the full result of our pp-test for every parameter and every dispersion case considered. As before, these are calculated using the KS test.

We obtain the lowest combined p-value (0.01) for $\alpha=2.5$ dispersion. It is driven primarily by the recovery of spin magnitudes ($a_{1,2}$), time of coalescence ($t_c$), and the angle between the total angular momentum and the line of sight ($\theta_{JN}$). This points to a possibility that our chosen number of live points ($n=1200$) for our injection test might have been too low to obtain fully accurate 16-D posteriors, at least for $\alpha = 2.5$ dispersion. For our test, we need only the marginalized $A_{\mathrm{eff}}$ posteriors though, and we have not found any problem with those in our injection tests. Still, while running the MDR test on real events, we have raised the number of live points to 1500.

\addtolength{\tabcolsep}{+2pt} 
\begin{table*}[]
    \centering
    \begin{tabular}{lllllllllllll}
 \toprule
  & $A_{-3}$ & $A_{-2}$ & $A_{-1}$ & $A_{0}$ & $A_{0.5}$ & $A_{1.5}$ & $A_{2.5}$ & $A_{3}$ & $A_{3.5}$ & $A_{4}$ & $m_g$ & $A_1$ \\
 \midrule
$\mathcal{M}_c$ & 0.75 & 0.43 & 0.21 & 0.95 & 0.77 & 0.09 & 0.70 & 0.65 & 0.88 & 0.91 & 0.23 & 0.20 \\
$q$ & 0.82 & 0.94 & 0.64 & 0.53 & 0.25 & 0.08 & 0.62 & 0.56 & 0.36 & 0.27 & 0.59 & 0.60 \\
$a_1$ & 0.42 & 0.26 & 0.16 & 0.25 & 0.05 & 0.81 & 0.00 & 0.07 & 0.02 & 0.20 & 0.81 & 0.51 \\
$a_2$ & 0.15 & 0.11 & 0.03 & 0.08 & 0.15 & 0.15 & 0.03 & 0.11 & 0.15 & 0.13 & 0.24 & 0.04 \\
$\theta_1$ & 0.74 & 0.51 & 0.82 & 0.73 & 0.38 & 0.31 & 0.45 & 0.75 & 0.60 & 0.85 & 0.65 & 0.54 \\
$\theta_2$ & 0.55 & 0.49 & 0.74 & 0.52 & 0.69 & 0.71 & 0.08 & 0.31 & 0.63 & 0.54 & 0.62 & 0.44 \\
$\phi_{12}$ & 0.51 & 0.86 & 0.85 & 0.91 & 0.85 & 0.55 & 0.95 & 0.92 & 0.88 & 0.99 & 0.67 & 0.07 \\
$\phi_{JL}$ & 0.19 & 0.45 & 0.29 & 0.11 & 0.36 & 0.15 & 0.48 & 0.67 & 0.48 & 0.10 & 0.44 & 0.48 \\
$D_L$ & 0.26 & 0.37 & 0.10 & 0.06 & 0.11 & 0.04 & 0.36 & 0.21 & 0.48 & 0.87 & 0.59 & 0.01 \\
$\mathrm{DEC}$ & 0.79 & 0.97 & 0.96 & 0.42 & 0.87 & 0.40 & 0.89 & 0.89 & 0.28 & 0.34 & 0.87 & 0.77 \\
$\mathrm{RA}$ & 0.51 & 0.33 & 0.82 & 0.99 & 0.57 & 0.10 & 0.42 & 0.51 & 0.84 & 0.46 & 0.30 & 0.79 \\
$\theta_{JN}$ & 0.19 & 0.07 & 0.41 & 0.05 & 0.01 & 0.25 & 0.03 & 0.23 & 0.09 & 0.05 & 0.27 & 0.17 \\
$\psi$ & 0.57 & 0.92 & 0.88 & 0.23 & 0.55 & 0.42 & 0.87 & 0.60 & 0.69 & 0.90 & 0.88 & 0.38 \\
$\phi$ & 0.33 & 0.89 & 0.64 & 0.69 & 0.33 & 0.27 & 0.25 & 0.82 & 0.91 & 0.40 & 0.48 & 0.54 \\
$t_c$ & 0.56 & 0.24 & 0.40 & 0.41 & 0.51 & 0.46 & 0.01 & 0.05 & 0.05 & 0.13 & 0.35 & 0.26 \\
$A_{eff}$ & 0.66 & 0.95 & 0.61 & 0.20 & 0.25 & 0.29 & 0.10 & 0.59 & 0.22 & 0.49 & 0.77 & 0.28 \\
\midrule
  & 0.76 & 0.73 & 0.62 & 0.22 & 0.13 & 0.05 & 0.01 & 0.47 & 0.26 & 0.38 & 0.90 & 0.09 \\
 \bottomrule
 \end{tabular}

    \caption{The results of the p-p test for all sampled parameters. For the dispersion case $A_1$ we sampled in the MDR phase, $\Psi_{MDR}$. The bottom row displays the combined p-value for a given injection set across all the parameters.}
    \label{tab:pp_values}
\end{table*}
\addtolength{\tabcolsep}{-2pt}

\FloatBarrier
\bibliography{bibliography}

\end{document}